\documentclass[twocolumn,tighten]{aastex61}

\usepackage{latexsym}
\usepackage{natbib}
\usepackage{amssymb}
\usepackage{amsmath}

\newcommand{\re}{$R_e$}

\newcommand{\muem}{$\langle\mu\rangle_e$}
\newcommand{\Ie}{$\langle I\rangle_e$}

\newcommand{\LnV}{$\log(L)-\log(n)$}
\newcommand{\RnV}{$\log(R_e)-\log(n)$}
\newcommand{\BmV}{$\rm{B-V}$}
\newcommand{\MV}{$M_V$}

\newcommand{\bfilt}{B}
\newcommand{\vfilt}{V}

\newcommand{\deV}{$r^{1/4}$}
\newcommand{\Sers}{$r^{1/n}$}

\newcommand{\ie}{{\em i.e.}}

\received{November 10, 2018}
\revised{November 10, 2018}
\accepted{November 10, 2018}
\submitjournal{ApJ}

\shorttitle{The parallelism between galaxy clusters and early-type galaxies: I.}
\shortauthors{D'Onofrio et al.}

\begin{document}

\title{The parallelism between galaxy clusters and early-type galaxies: I. \\
The light and mass profiles}

\correspondingauthor{Mauro D'Onofrio}
\email{mauro.donofrio@unipd.it}

\author[0000-0001-6441-9044]{Mauro D'Onofrio}
\affil{Department of Physics and Astronomy, University of Padova, Vicolo Osservatorio 3, I35122, Italy}

\author{Mauro Sciarratta}
\affil{Department of Physics and Astronomy, University of Padova, Vicolo Osservatorio 3, I35122, Italy}

\author{Stefano Cariddi}
\affil{Department of Physics and Astronomy, University of Padova, Vicolo Osservatorio 3, I35122, Italy}

\author{Paola Marziani}
\affil{INAF - Astronomical Observatory of Padova, Vicolo Osservatorio 5, I35122, Italy}

\author{Cesare Chiosi}
\affil{Department of Physics and Astronomy, University of Padova, Vicolo Osservatorio 3, I35122, Italy}

\begin{abstract}

We have analyzed the parallelism between the properties of galaxy clusters and early-type galaxies (ETGs) by looking at the similarity between their light profiles. We find that the equivalent luminosity profiles of all these systems in the \vfilt\ band, once normalized to the effective radius \re\ and shifted in surface brightness, can be fitted by the S\'ersic's law \Sers\ and superposed with a small scatter ($\le0.3$ mag). By grouping objects in different classes of luminosity, the average profile of each class slightly deviates from the other only in the inner and outer regions (outside $0.1\leq r/R_e\leq 3$), but the range of values of $n$ remains ample for the members of each class, indicating that objects with similar luminosity have quite different shapes. 

The "Illustris" simulation reproduces quite well the luminosity profiles of ETGs, with the exception of in the inner and outer regions where feedback from supernovae and active galactic nuclei, wet and dry mergers, are at work. The total mass and luminosity of galaxy clusters as well as their light profiles are not well reproduced.

By exploiting simulations we have followed the variation of the effective half-light and half-mass radius of ETGs up to $z=0.8$, noting that progenitors are not necessarily smaller in size than current objects. We have also analyzed the projected dark+baryonic and dark-only mass profiles discovering that after a normalization to the half-mass radius, they can be well superposed and fitted by the S\'ersic's law.

\end{abstract}

\keywords{Clusters of galaxies -- Early-type galaxies -- structure -- photometry -- scaling relations -- simulations}

\section{Introduction} \label{sec:intro}

\noindent
Recently \cite{Cariddietal2018} derived the equivalent luminosity profiles in the \vfilt\ and \bfilt\ bands of several nearby galaxy clusters observed by the survey WINGS and Omega-WINGS
\citep{Fasano2006,Varela2009,Moretti2014,Donofrio2014,Gullieuszik2015}. In these light profiles the average surface brightness is measured on circles of increasing radius centered on the position of the brightest cluster galaxy.
Their work showed that galaxy clusters share many properties in common with early-type galaxies (ETGs): the behavior of the growth curves and of the surface brightness profiles, the distribution in the Kormendy and Faber-Jackson relations \citep{Kormendy1977,FaberJackson} and the existence of a similar color-magnitude diagram. They further showed that galaxy clusters are best fitted by the S\'ersic \Sers\ law \citep{Sersic1968} and might be non-homologous systems like ETGs \citep{Caonetal1993,Donofrio94}.

These striking parallelisms between systems so different in size are particularly interesting when the mechanisms of structure assembly are considered in the current cosmological framework. These parallelisms are not unknown: previous works already noted some interesting similarity between clusters and ETGs. The most known example is the observed distribution of clusters in the Fundamental Plane (FP) relation\footnote{The Fundamental Plane is the relation between the effective radius \re, the effective surface brightness \muem\ and the central velocity dispersion $\sigma_0$.} \citep{Dressleretal1987,Djorgovski&Davis1987}.
\cite{Schaeffer}, \cite{Adami} and \cite{Donofrio2013} showed that galaxy clusters share approximately the same FP relation of ETGs. This is not surprising if we think that clusters are almost virialized structures with a similar dynamics. The FP relation is in fact a universal relation for several gravitating systems closed to the virial equilibrium, such as GCs \citep[see, e.g.][]{Djorgovski1995,McLaughlin2000, Barmbyetal2007}, cluster spheroids \citep{Zaritskyetal2006}, open clusters \citep{Bonatto&Bica2005}, X-ray emitting elliptical galaxies \citep{Diehl&Statler2005}, clusters of galaxies \citep[see e.g.][]{Lanzonietal2004}, supermassive black holes (BHs) \citep[e.g.][]{Hopkinsetal2007} and quasars \citep[e.g.][]{Hamiltonetal2006}. Its tilt with respect to the prediction of the Virial theorem has several possible origins.

Another example comes from the luminosity profiles. ETGs and clusters are best fitted by the S\'ersic \Sers\ law \citep[see e.g.][]{Caonetal1993,Cariddietal2018} and it is well known that the index $n$ correlates with many observable parameters, such as luminosity \MV,  effective radius \re\ and effective surface brightness \muem\ \citep{Caonetal1993,GrahamGuzman2003}. The systematic variation of $n$ in ETGs is interpreted as a deviation from structural and dynamical homology in collisionless stellar systems \citep{Ciotti1991,CiottiBertin1999,Graham2001,Graham2002,Grahametal2003,Trujilloetal2004}. The final value of $n$ probably results from several dissipationless merging events. The nearly exponential value of $n$ ($\sim1$) observed in dwarf systems is similar to that of disk objects, suggesting that gas collapsed and dissipated energy, while the large value of $n$ in luminous ETGs is due to merging, with wet events in the remote epochs and dry events in the recent ones. This scenario seems supported by numerical simulations \citep[see e.g.][]{ScannapiecoTissera2003,Eliche-Moraletal2006}.
If clusters are indeed non-homologous systems like ETGs, what produces the observed variations of $n$? Why clusters share the same properties of ETGs when we look at their luminosity profiles? 

\cite{Cariddietal2018} further showed that in the \MV\ $-$(\BmV) plane galaxy clusters share the same "red-sequence" slope of ETGs, when the mean color is measured within $0.6R_{200}$\footnote{$R_{200}$ is the radius at which the projected density is 200 times the critical density of the Universe.}. In other words small/faint clusters are bluer than big/bright clusters. The same trend is observed in galaxies. Why we see such a similar color-magnitude diagram?

On the theoretical side numerical simulations of the cold dark matter (CDM) universe predict that halos are nearly self-similar \citep[see e.g.]{Dubinski&Carlberg1991,NFW1996,Mooreetal1998,Ghignaetal2000}. 
The collapsed halos have in general a central density cusp with a $\rho\propto r^{-1}$ profile that at large radii becomes $\rho\propto r^{-3}$. These slopes are those characteristic of the Navarro-Frenk-White (NFW) profile \citep{NFW1996}, which provides a good description of N-body simulations.
\cite{Merrittetal2005} however showed that the S\'ersic law might provide equally good fits for the density profiles of dark matter (DM) halos with small central deviations that can be attributed to dynamical effects \citep{vanderMarel1999,Milosavljevicetal2002,RavindranathHoFilippenko2002,Graham2004,Merrittetal2004,PretoMerrittSpurzem2004}. This suggests that a "universal" mass profile might exist for galaxies and clusters and that the observed deviations originate in the complex physics of the luminous baryon component.

Improved simulations at high resolution showed that DM halos are not strictly self-similar \citep{Merrittetal2006,Navarroetal2010} and can also vary their shape for the action of baryonic matter. Cooling might allow baryons to condense toward the center, producing an higher concentration of DM \citep[e.g.]{Blumenthaletal1986,Gnedinetal2004,Pedrosaetal2009,Abadietal2010} and heating of the central cusp could be induced by dynamical friction \citep[e.g][]{Elzantetal2001,Nipotietal2004,RomanoDiazetal2008,DelPopolo&Cardone2012}, feedback from supernovae  \citep{Governatoetal2010,Governatoetal2012}, and feedback from active galactic nuclei \citep{Peiranietal2008,Martizzietal2012}. 

Understanding the relative contribution of the dark and luminous components in ETGs and clusters is therefore important. As a matter of fact if the CDM description of the universe is correct, then the structure of real halos of such systems can provide several information about the assembly of the structures themselves indicating what was the imprint of baryons on their halos. The dark and baryonic density profiles and the distribution in the main scaling relations can furthermore inform us about the relative importance of dissipational and dissipationless processes occurred in the evolution of systems as well as on the relative weight of feedback processes and assembly of the central super massive black-holes (SMBHs). 

This is the first paper of a sequel dedicated to the study of clusters and ETGs. The aim is to describe the properties that these systems have in common and to test the ability of current hydrodynamical models in reproducing the observational evidences.
The paper is designed as follows: in Sec. \ref{sec:1} we introduce the galaxy sample and the construction of the equivalent surface brightness profiles of galaxies and clusters; in Sec. \ref{sec:2} we present the results obtained on real galaxies,
comparing the light profiles of ETGs and clusters (Sec. \ref{sec:2:1}) and discussing the non-homology of ETGs and clusters (Sec. \ref{sec:2:2}); in Sec. \ref{sec:3} we introduce the Illustris dataset of hydrodynamical simulations presenting the light profiles derived for galaxies and clusters \citep{Vogel2014,Genel_etal_2014,Nelsonetal2015}; in Sec. \ref{SecSim} we check the degree of non-homology emerging from simulated data, in Sec. \ref{Progenitors} we look at the evolution of the effective radius of ETGs up to redshift $z=0.8$ and in Sec. \ref{sec:surfacemass} we analyze the mass profiles of ETGs and clusters coming from the simulated data; finally, in Sec. \ref{sec:4} we summarize our conclusions. In Appendix we provide several tables with the data used in this work.

Throughout the paper we assumed in all our calculations the same values of the $\Lambda$-CDM cosmology used by the Illustris simulation \citep{Vogel2014}: $\Omega_m = 0.2726$, $\Omega_{\Lambda}= 0.7274$, $\Omega_b = 0.0456$, $\sigma_8 = 0.809$, $n_s = 0.963$, $H_0 = 70.4$ km s$^{-1}$ Mpc$^{-1}$. 

\section{The sample} \label{sec:1}

The sample of clusters analyzed here is that of the WINGS and Omega-WINGS surveys \citep{Fasano2006,Varela2009,Moretti2014,Donofrio2014,Gullieuszik2015}. With respect to the original WINGS sample, the number of clusters is limited to 45 objects \citep[those of the southern hemisphere covered by the Omega-WINGS survey with available spectroscopic measurements;][]{Cava2009,Morettietal2017}.
The luminosity profiles of these galaxy clusters were derived by \cite{Cariddietal2018} by integrating the luminosity growth curves in the \vfilt\ and \bfilt\ bands.  These were obtained by statistically subtracting the contribution of background objects and taking into account all sources of incompleteness. The growth curves were then fitted using the S\'ersic's law and transformed in surface brightness units. 
The corresponding clusters structural parameters, the half light radius \re\ and the mean surface brightness \Ie\ were also derived by \cite{Cariddietal2018} by integrating the circular luminosity growth curves of clusters. 

\begin{figure*}
\includegraphics[scale=0.9,angle=-90]{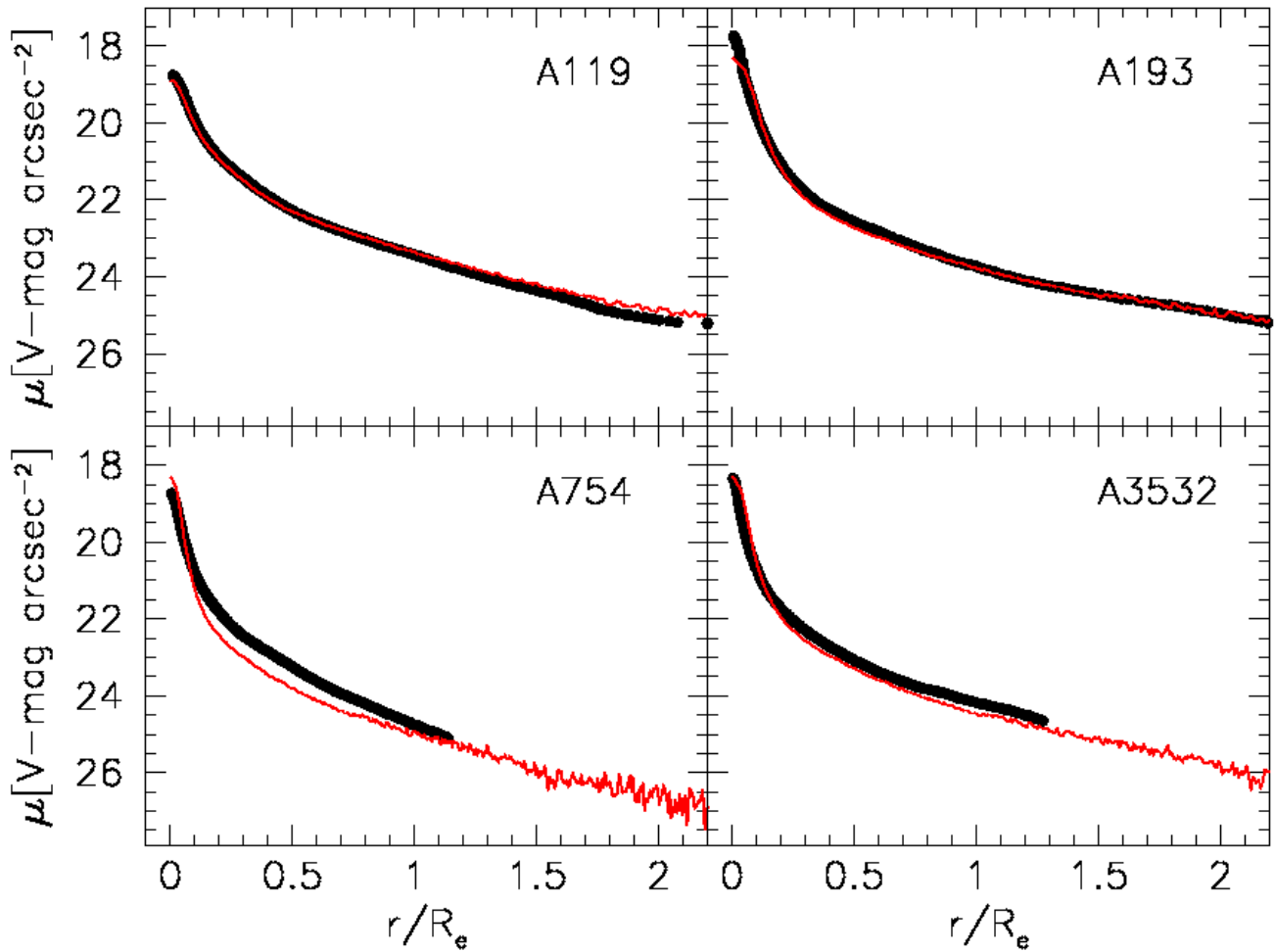}
	\caption{Comparison between the equivalent surface brightness profiles of the BCGs derived with GASPHOT (black lines) and AIAP (red lines). The cluster name is labeled in each box.}
		\label{fig:bestmatchBCG}
\end{figure*}

\begin{figure}
\begin{center}
\includegraphics[scale=0.42,angle=0]{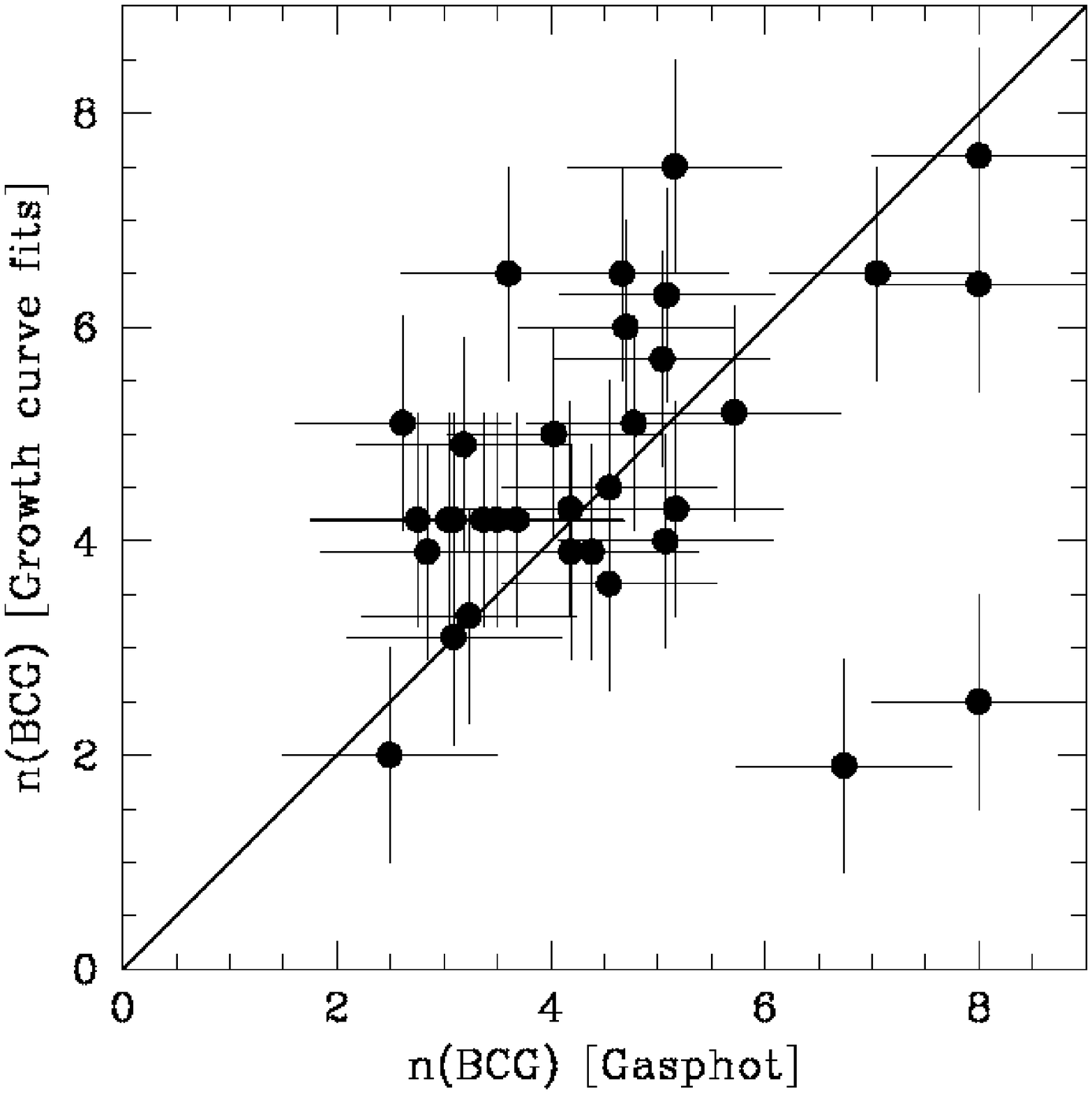}
	\caption{Comparison of the S\'ersic indexes of the BCGs derived with GASPHOT and those derived from the fit of the growth curves. The errors on $n$ have been set to 1 for all objects (see text).}
	\label{fig:Sersic-comp}
	\end{center}
\end{figure} 

The ETGs studied here for the comparison with galaxy clusters are the brightest (BCG) and second brightest (II-BCG) cluster galaxies plus a sample of normal ETGs members of the clusters (one for each cluster) randomly chosen in the CCD images. 
The main structural parameters of these galaxies were already derived by \cite{Donofrio2014} through the software GASPHOT \citep{Pignatelli}. However,
for these objects we have re-derived here the equivalent surface brightness profiles in a completely independent way using the software AIAP \citep{Fasanoetal2010}.
The reason of this re-analysis is that we want to compare in a rigorous way the light profiles of galaxies and clusters up to the faintest level in surface brightness. 
Fig.\,\ref{fig:bestmatchBCG} shows an example of the accuracy reached by both methods of analysis. The GASPHOT and AIAP profiles are compared for four objects taken randomly from the sample. Note that the GASPHOT profiles are less deep in surface brightness and are less noisy.
Some differences among the profiles are present and are due to the fact that the two software packages AIAP and GASPHOT work in a different way. AIAP performs the photometric analysis of the single galaxies by constructing manually all the isophotes. All sources of disturbance (stars, background galaxies, stellar spikes, etc.) are previously masked and the data are smoothed when the sky background does not permit to draw a clear isophote. All isophotes are then fitted by ellipses and fitted with the \deV\ or \Sers\ law. On the other hand GASPHOT is an automatic procedure based on the Sextractor analysis. This software does not permit the accuracy of the manual procedure, in particular when disturbing objects are superposed to the galaxy image. The fainter surface brightness reached by AIAP is a consequence of this different approach.

The typical photometric error in the light profile is around $\sim 10\%$ at $\mu(V)\sim26$ mag arcsec$^{-2}$ \citep{Fasano2006}.
The photometric error bars are not shown in the figure. They are in fact poorly known for several reasons: 1. the automatic procedure performs the fit of the major and minor axes profiles of the galaxies by convolving the \Sers\ law with an average point spread function (PSF) derived for the image. It follows that the photometric error close to the central region might be larger than $10\%$; 2. in the outer region the effects of the sky subtraction can be substantial. The software subtracts a constant value derived in a region around each galaxy, but in some cases this is not sufficient to get an high accuracy in the profiles. The sky background might change for the proximity of bright objects. Since an accurate determination of the errors in the profiles would require a much complex analysis and our aim here is to compare the average profiles of classes of objects we have not taken into account the errors of the single profiles, but only the standard deviation around the mean of the average profiles.

The S\'ersic index $n$ of our objects has been also re-measured by fitting the AIAP growth curve profiles. Fig.\,\ref{fig:Sersic-comp} shows the comparison between the S\'ersic indexes measured by GASPHOT and AIAP. We see that there is a substantial agreement, but the typical error $\Delta n$ is quite large ($\sim 1$). The large uncertainty in the value of $n$ depends on several factors: the interval used for the fit, the FWHM of the seeing, the correct estimate of the sky background in the surface brightness profiles and in the growth curve profiles. For these reasons we decided to set the error $\Delta n$ to 1 for all the data and we have only shown the equality line. The aim of Fig.\,\ref{fig:Sersic-comp} is simply to stress the difficulty of trusting in the values of $n$ when different methods are used to get it.
At the end of this re-analysis we decided to keep the structural parameters of galaxies derived by GASPHOT, since this software provides a seeing convolved fit of the major and minor axes surface brightness profiles of each object at the same time, giving in output the S\'ersic index $n$, the effective radius \re, the effective surface brightness \Ie\ and the average flattening $\langle b/a\rangle$. The AIAP profiles have been instead used in our figures to better show the behavior of the profiles at the faintest level of surface brightness.

\section{Results with the observational data}\label{sec:2}
We start here the presentation of the results obtained from our analysis of the observational data.

\subsection{The light profiles of the ETGs and clusters} \label{sec:2:1}

First we remind that the profiles are in the $V$ band and are equivalent profiles, \ie\ average surface brightness measured in circles of increasing radius centred on the peak luminosity. All profiles have been normalized to the effective radius, that enclosing half the total luminosity.

What we discuss is not the single profile of each object, but the average profile for the class of objects defined in the above section (BCGs, II-BCGs, random normal ETGs and clusters). The average profiles have been derived in two ways: 1) by considering the mean value of the surface brightness for all the profiles of each class, and 2) by shifting all
the profiles with respect to one reference object.  In this case, after finding the shift that produces the lowest average residuals in the selected interval, we have derived the mean profile among the whole set of shifted profiles.

\begin{figure}
\begin{center}
\includegraphics[scale=0.42,angle=0]{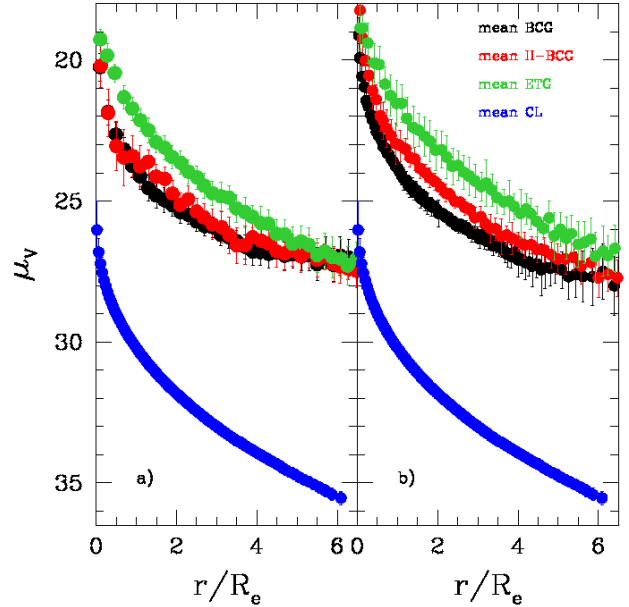}
	\caption{Panel a): the average surface brightness profiles for each class of objects (galaxies and clusters). The average profile of BCGs is marked by black dots, II-BCGs by red dots, normal ETGs by green dots and clusters by blue dots. The error bars give the $1\sigma$ standard deviation around the mean value. Panel b): in this plot all the profiles have been previously shifted with respect to the corresponding reference object of A160 and then averaged (see text).}
	\label{fig:Allprof}
	\end{center}
\end{figure}

We have chosen the BCG, II-BCG, and the normal ETG in A160 as reference for galaxies and the profile of the cluster A160 as reference for clusters. The choice of A160 as reference does not influence the conclusions we draw below. Any other reference object in different clusters produces the same results.  The two methods do not produce significative different results. The left panel of Fig.\,\ref{fig:Allprof} shows the average profiles obtained for BCGs (black dots), II-BCGs (red dots) and normal ETGs (green dots). Note the small difference with the right panel obtained with the second method. The average profiles indicate
that normal ETGs are systematically brighter than BCGs in surface brightness at each $r/R_e$. II-BCGs are also a bit brighter than BCGs.
We will see in Sec. \ref{sec:3} that this behavior is reproduced by numerical simulations.

The average cluster profile shown in blue color in Fig.\,\ref{fig:Allprof} was obtained by excluding the profiles of the following clusters: A85, A147, A151, A168, A1991, A2399, A2415, A2457, A3158, A3528a, A3809, A3128, A3880. The profiles of these clusters show clear evidences of the presence of more complex structures (double components and anomalous behavior of the growth curves), that we believe are likely originated by recent merging events occurred in these clusters. The shape of these profiles indicates that in our sample $\sim 30\%$ of the clusters have not yet reached a completely relaxed photometric structure (i.e. the profiles cannot be fitted with one single S\'ersic component). This is a significant fraction; however the inclusion of non S\'ersic profiles would not permit a correct comparison with ETGs.

Fig.\,\ref{fig:Allprof} clearly shows that the surface brightness profiles of each class, once normalized to the effective radius, can be superposed with a small scatter (the typical rms is $\sim0.3\div0.5$ mag) once a constant shift in surface brightness is applied. 
To obtain a superposition of the profiles in surface brightness, depending on the interval chosen to realize the match, one must
add one of the values reported in Tab. \ref{tab:shift_gal}.

\begin{figure}
\includegraphics[scale=0.4,angle=0]{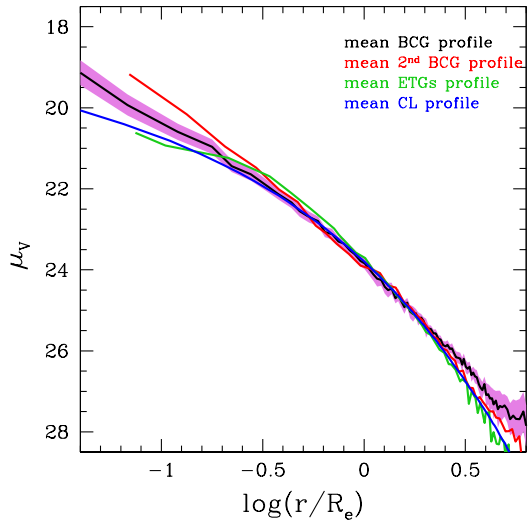}
	\caption{The average surface brightness profiles for each class of objects in log scale. The color code is in the legend. The equivalent surface brightness profiles are shown here superposed with respect to the BCG profile chosen as reference. The shaded area marks the $1\sigma$ error in the distribution of the BCG profiles around the reference BCG in A160.}
	\label{fig:Allprof2}
\end{figure}

The result of the superposition is better visible in Fig.\,\ref{fig:Allprof2} where the profiles of all classes have been shifted by a constant value using the average profile of the BCGs as reference. 
We have imposed a vertical shift minimizing the difference
in $\mu_V$ in all bins of $r/R_e$: the shift is calculated as:
\begin{equation}\label{eqshift}
    s_j = \frac{1}{N} \sum_{i=1}^{N} \mu_{simBCG}(r_i/R_e) - \mu_{j}(r_i/R_e)
\end{equation}
where $N$ is the number of radial bins in which the difference is evaluated, while $j$ refers to any other kind of structures except simulated BCGs. All values of $s_j$ are reported in Table \ref{tab:shift_gal}.

\begin{table}[]
    \centering
    \caption{Values of the constant shift $s_j$ used to superpose the average surface brightness profiles in
    Fig.\,\ref{fig:Allprof}. The average profile used as reference is that of BCGs.}
    \renewcommand{\arraystretch}{0.8}
    \begin{tabular}{|c|c|}
    \hline
    Obs. obj. & $s_j$ \\
    \hline
    BCG    & 0.0 \\
    II-BCG & 0.4$\div$0.8 \\
    ETG   & 1.5$\div$2.0 \\
    CL     & -6.5 $\div$ -7.0 \\
    \hline
    \end{tabular}
    \label{tab:shift_gal}
\end{table}

The profiles in Fig.\,\ref{fig:Allprof2} are shown in log scale to better put in evidence the differences among them.
The figure indicates that all profiles are well superposed (within $\sim 0.3$ mag) in the interval $0.3\leq r/$\re$\leq 2.5$, while the differences emerge in the center ($r/$\re$\leq 0.3$) and in the outer regions ($2.5\leq r/$\re$\leq 6.3$).

\begin{figure*}
    \centering
	\gridline{\fig{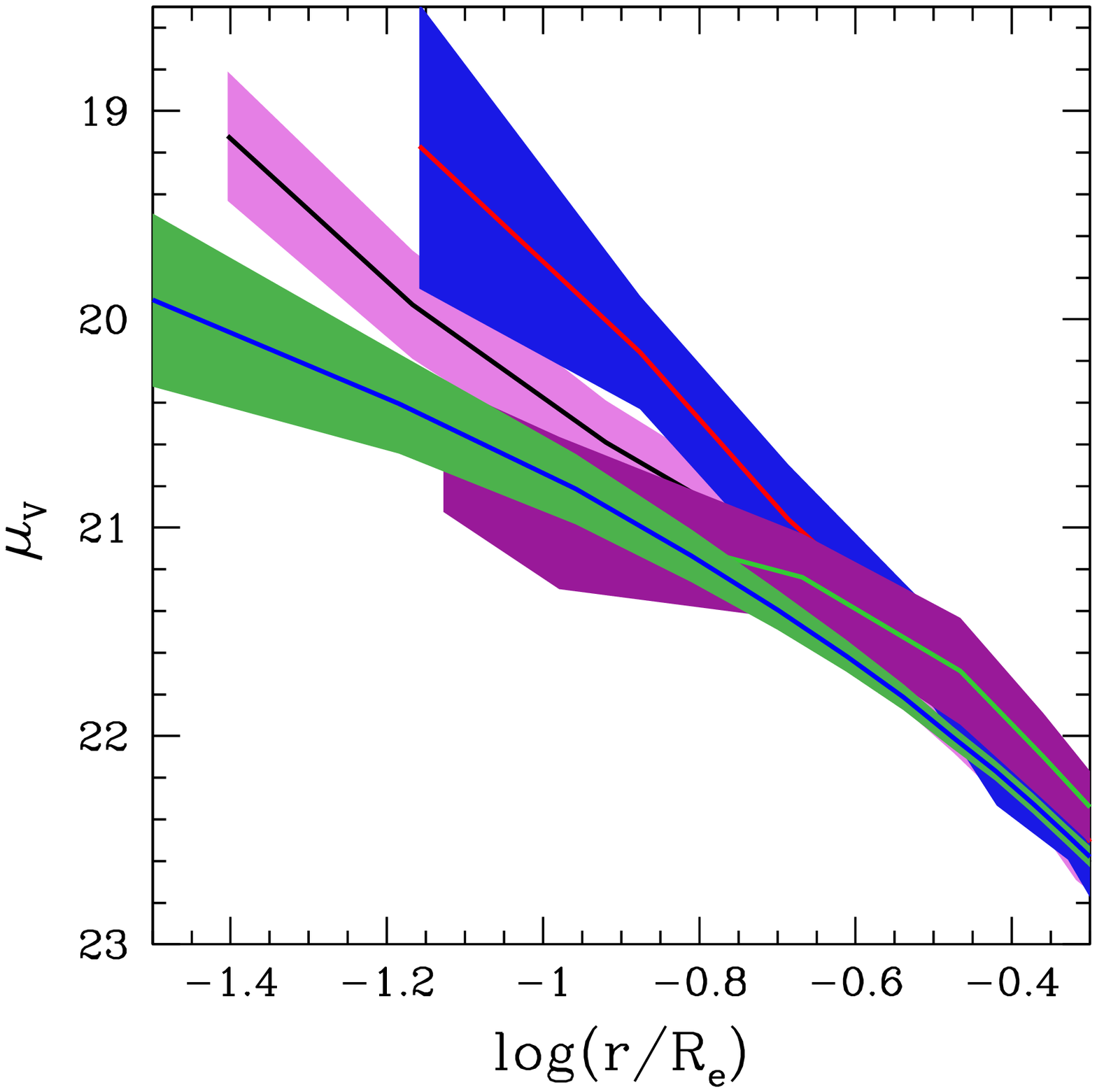}{0.5\textwidth}{}
		\fig{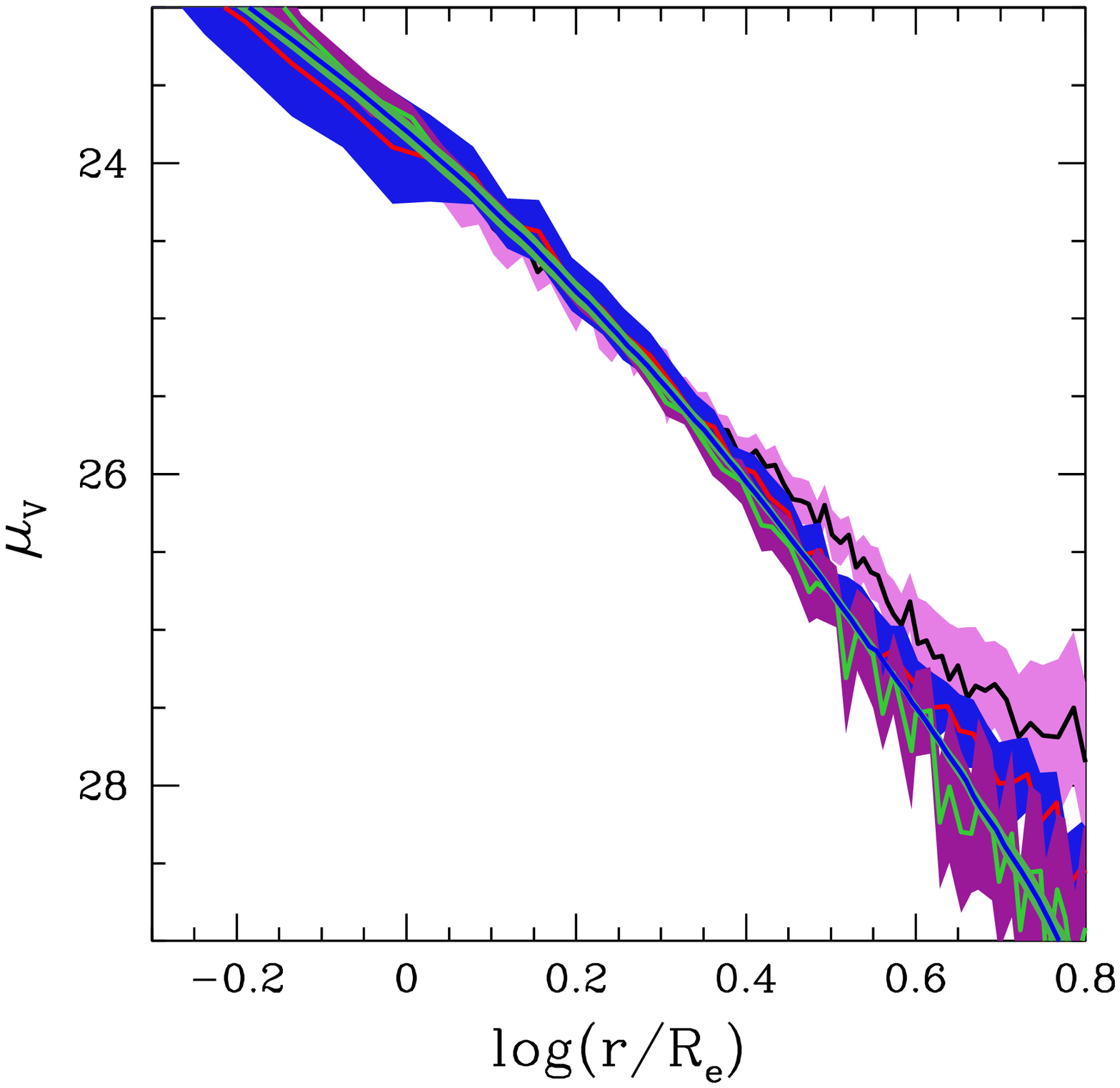}{0.5\textwidth}{}
	}
    \caption{Zoom on the inner region (left panel) and outer region (right panel) of Fig.\,\ref{fig:Allprof2}. The colored bands mark the $1\sigma$ error around the mean profile of each class. As in previous figures the black line marks BCGs, the red II-BCGs, the green normal ETGs and the blue clusters. The shaded area in pink color is used for BCGs, the blue for II-BCGs, the violet for normal ETGs and the green for clusters.}
    \label{fig:Innprof}
\end{figure*}

Since the seeing was not exactly the same in each CCD image, the average profile we get is still affected by seeing in its central part. However, {to a first order approximation} this is not strictly a problem for our analysis since we are interested in the relative comparison between the profiles of different classes. After averaging the profiles, the same average seeing effect is present in all profiles, so that the relative difference can be considered almost free from the effect of seeing. The average cluster profile is not affected by seeing being simply the sum of the light of the single galaxies members of clusters.

The differences observed in the central region of Fig.\,\ref{fig:Allprof2} require however a much careful analysis of the seeing effects. The seeing effects can be quite different for small and big galaxies. Furthermore one should also consider that all profiles are re-scaled in units of $r/R_e$, with \re\ spanning a big interval of values (see below).

We have therefore repeated the analysis of the profiles using the S\'ersic growth curve models obtained by GASPHOT. These models once convolved with the PSF give the observed growth curves that divided by the area provide the surface brightness profiles. They are in practice the deconvolved integrated magnitude profiles of our objects, not affected by the seeing. 

Fig.\,\ref{fig:Innprof} shows a zoom of the inner and outer regions of Fig.\,\ref{fig:Allprof2}. The colored bands mark the $1\sigma$ error around the mean profile of each class. We clearly see that in the inner region the mean profile of BCGs differ from that of II-BCGs and clusters at more than $1\sigma$, while normal ETGs are always consistent with the cluster profile. In the outer region such differences remain although less marked.

\begin{figure}
\includegraphics[scale=0.4,angle=0]{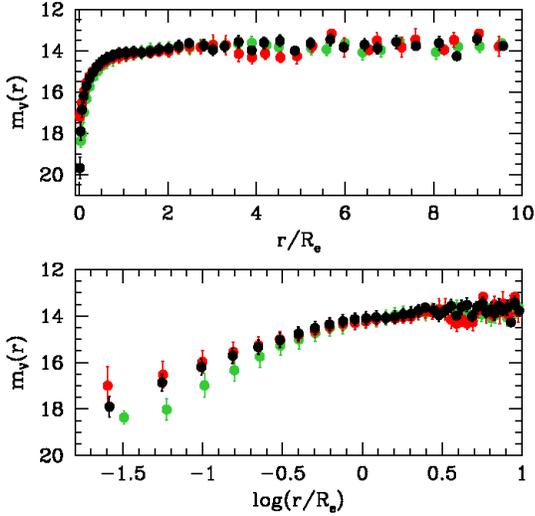}
	\caption{Top panel: The average deconvolved growth curve profiles for each class of objects in linear scale. Radii are normalized to the effective radius. As before BCGs are marked by black dots, II-BCGs by red dots and normal ETGs by green dots. The errobars give the $1\sigma$ standard deviation around the mean profile.}
	\label{fig:Allprof_dec}
\end{figure}

Fig.\,\ref{fig:Allprof_dec} shows instead in linear (top panel) and log scale (bottom panel) the growth curves of our classes of objects shifted artificially to the BCG curve. Note that in this figure we have the integrated magnitude within circular apertures and not the surface brightness. The log scale indicate that normal ETGs once superposed to BCGs are in general fainter at each $r/R_e$, while II-BCGs are brighter than BCGs. This confirms that the seeing effects have not significantly affected the distribution observed in Fig.\,\ref{fig:Allprof2}.

\begin{figure}
\includegraphics[scale=0.4,angle=0]{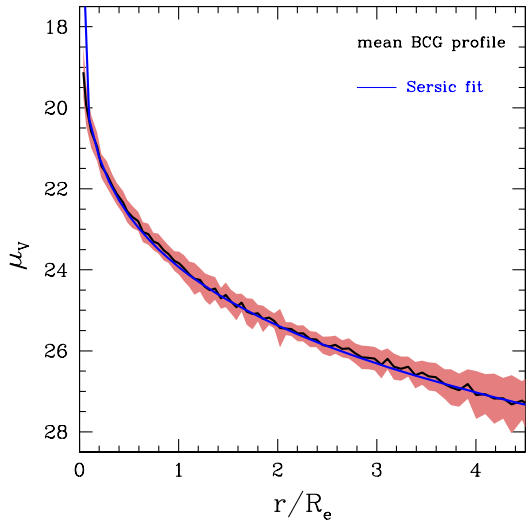}
	\caption{The fit of the average profile of BCGs (black line) with the S\'ersic law (blue line).}
	\label{fig:SersicfitBCG}
\end{figure}

\smallskip
Fig.\,\ref{fig:SersicfitBCG} shows the average profile of BCGs and the fit we get with the S\'ersic's law. In general good fits are obtained for all averaged profiles with this empirical law. 

\begin{table}[]
    %\centering
    \caption{Best S\'ersic fit parameters for the average surface brightness profiles of observed and simulated galaxies and clusters. The errors related to the parameters and the values of the $1\sigma$ scatter are reported. The fits have been computed in the range $r/R_e \in [0,6]$ for real objects and in the range $r/R_e \in [0,3]$ for simulated objects.}
    \begin{tabular}{|c|c c c c c|}
    \hline
Obs. obj. (z=0)  & $n$ & $\Delta n$ & $b(n)$ & $\Delta b(n)$ & $\sigma$ \\
    \hline
    BCGs & 5.27 & 0.28 & 9.46 & 0.53 & 0.08 \\
    II-BCGs & 4.76 & 0.39 & 9.40 & 0.84 & 0.11 \\
    ETGs & 2.78 & 0.13 & 5.56 & 0.30 & 0.18 \\
    CLs & 2.58 & 0.01 & 4.81 & 0.02 & 0.01 \\
    \hline
        \hline
Sim. obj. (z=0)  & $n$ & $\Delta n$ & $b(n)$ & $\Delta b(n)$ & $\sigma$ \\
    \hline
    BCGs & 6.41 & 0.31 & 12.52 & 0.65 & 0.05\\
    II-BCGs & 4.93 & 0.15 & 9.72 & 0.32 & 0.05 \\
    ETGs & 1.77 & 0.03 & 3.17 & 0.08 & 0.19 \\
    CLs & 3.15 & 1.35 & 5.60 & 2.50 & 0.23 \\
   \hline
    \end{tabular}
    \label{Tabfitprofiles}
\end{table}

The fitted equation is:
\begin{equation}
\mu(r)=\mu_e+\frac{2.5b_n}{ln(10)}[(\frac{r}{R_e})^{1/n}-1].
\label{eqfit}
\end{equation}

Table \ref{Tabfitprofiles} gives the S\'ersic index $n$ and the value of the term $b_n$ from eq. \ref{eqfit} together with their uncertainty. The last column reports the $1\sigma$ scatter around the fit.
Note how the value of $n$ for galaxies decreases when the fit is done for objects of small mass.
The $1\sigma$ error around the fits is in general very low, indicating that the S\'ersic's law works very well either for ETGs and clusters. 

The good match between the light profiles of clusters and ETGs and the good fits obtained with the S\'ersic's law for both classes are the first element of the claimed parallelism.

\subsection{Non homology} \label{sec:2:2}

The non-homology of ETGs is well known after \cite{Caonetal1993,Donofrio94,GrahamColless}. They showed that the surface brightness profiles of ETGs are best fitted by the S\'ersic's law \Sers\ and the index $n$ is correlated with the effective radius and the total luminosity. This means that the average shape of ETGs depends on the total mass (and luminosity) of the systems. Some years ago ETGs were thought perfectly homologous systems, i.e. self-similar structures simply scaled for a constant factor, with luminosity profiles following the de Vaucoulers \deV\ law.

\begin{figure*}
    \centering
	\gridline{\fig{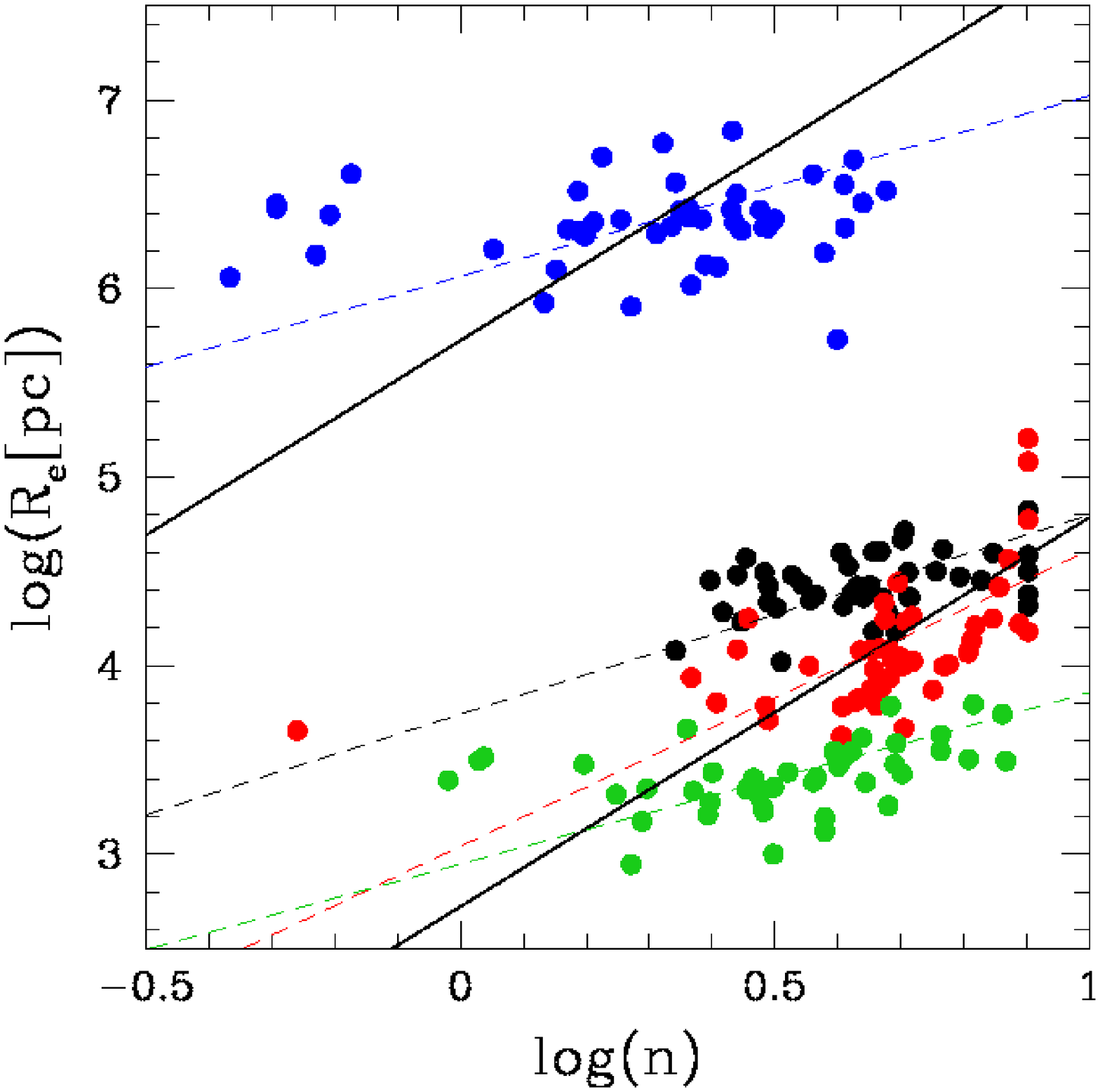}{0.5\textwidth}{}
		\fig{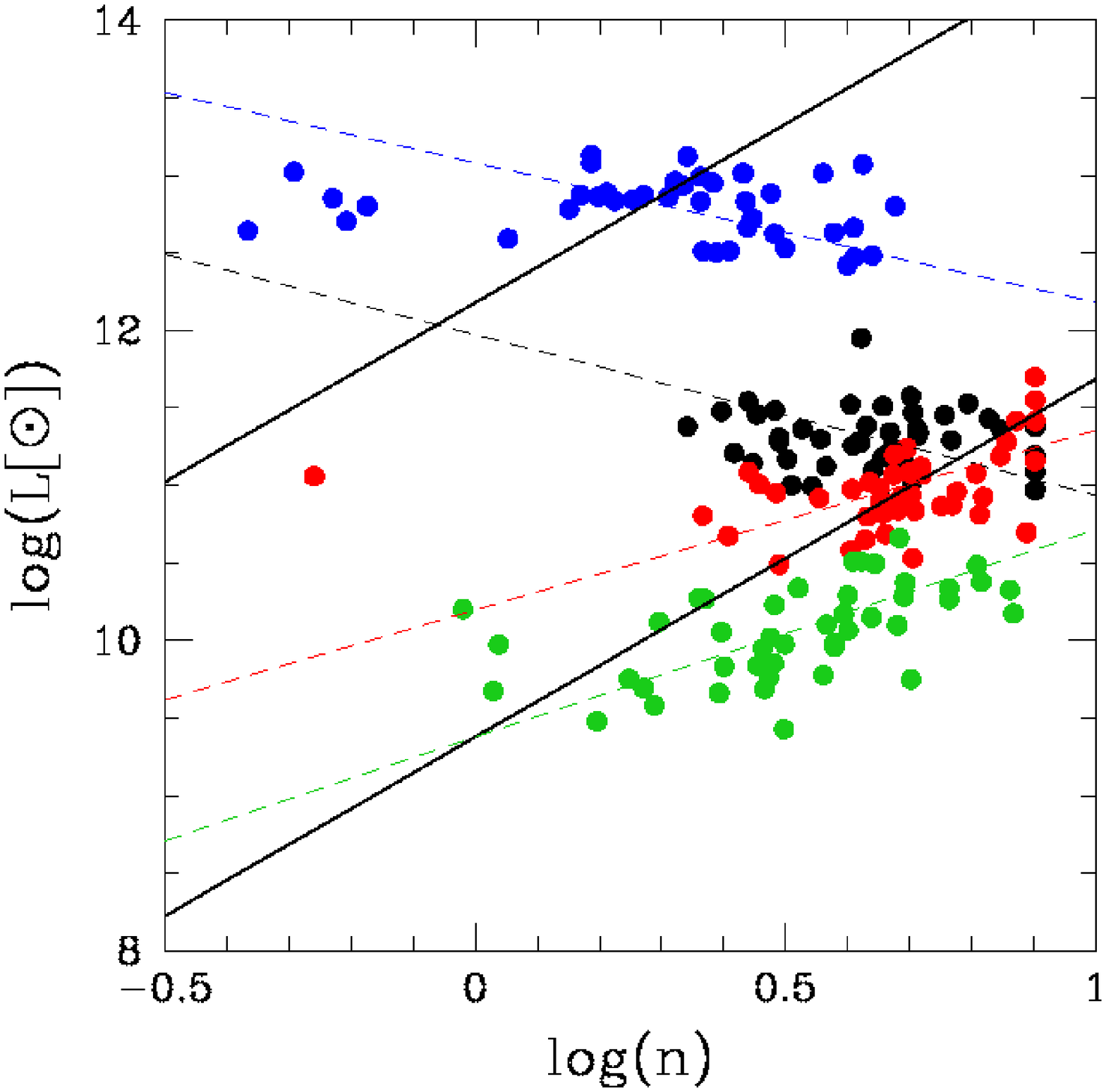}{0.5\textwidth}{}
	}
    \caption{Left panel: Distribution of ETGs and clusters in the \RnV\ plane. Black dots are BCGs, red dots are II-BCGs, green dots are random ETGs and blue dots are clusters. The bottom black solid line marks the fit of the whole distribution of ETGs (BCGs+II-BCGs+random normal ETGs). The upper solid line is a shifted version of this fit used to best match the cluster distribution. Right panel: Distribution of ETGs and clusters in the \LnV\ plane. The symbols are the same of the left panel. The bottom black solid line marks the fit of the whole distribution of ETGs (BCGs+II-BCGs+random normal ETGS). The upper solid line is a shifted version of the bottom solid line done to best match the cluster distribution. The thin colored dashed lines are the individual fits obtained by the program SLOPES (see text) for the different classes of objects.
    }
    \label{fig:nR}
\end{figure*}

The left panel of Fig.\,\ref{fig:nR} shows the distribution of our systems in the \RnV\ plane.  The values of $n$ used here are those derived by \cite{Donofrio2014} with GASPHOT for the singles ETGs and those of \cite{Cariddietal2018} for clusters. Note the solid line which represents the best fit of the whole distribution of ETGs (BCGs+II-BCGs+random ETGs). This line has been shifted along the y-axis matching
the distribution of clusters. We have chosen the shift that minimize the average difference of the residuals. The fit has been obtained with the standard lsq technique using the program SLOPES \cite{Feigelson1992}, which performs a bootstrap and a jackknife resampling of the data, without taking into account the measurement errors. We get: $\log(R_e)=2.0\log(n)+2.72$ with a rms$=0.42$, a c.c. of 0.48 and a Spearman significance of $2.0\cdot 10^{-9}$. This means that the correlation exists and is quite robust. Note that, looking at the single families of objects of different colors in the the plot, the correlation between the two variables is much less pronounced and even absent for clusters, although the interval of values of $n$ is quite large for each class. The global fit clearly does not represent the trend observed for each class of objects taken separately (see the dashed thin lines). A smaller slope is obtained for each class: 1.06 (BCGs), 1.57 (II-BCGs), 0.91 (random normal ETGs), 0.96 (clusters). The rms/c.c. is 0.15/0.38 (BCGs), 0.27/0.55 (II-BCGs), and 0.17/0.37 (random normal ETGs), 0.22/0.12 (clusters). The resampling analysis gives an error on the slopes of 0.2 for the fit of the whole ETG sample, so that in practice the observed difference with the other slopes is statistically significant.

This behavior is expected because we have defined three categories of objects with well defined values for the mass and luminosity in a quite restricted interval of values (see the top panel of Fig.\,\ref{fig:magnitudes}). Observe that in each class the shape is far from being similar for all objects. The values of $n$ are quite large in each class, so that the correlation with \re\ almost disappears.

For clusters the significance of the correlation is small ($4.2\cdot 10^{-1}$), while it is always better than $10^{-3}$ for all the other systems. The spread of values of $n$ however is quite large.

Along the same line the right panel of Fig.\,\ref{fig:nR} shows the distribution of our systems in the \LnV\ space.
Here again the fit is quite good for the whole distribution of ETGs:
$\log(L)=2.31\log(n)+9.38$ with a rms$=0.53$, a c.c. of 0.42 and a significance  of $1.5\cdot 10^{-7}$. For the single classes of objects we get for the slope/rms/c.c.: -1.03/0.19/-0.065 (BCGs), 1.16/0.23/0.32 (II-BCGs), 1.34/0.25/0.52 (random normal ETGs), -0.09/0.19/-0.19 (clusters). The error on the slope obtained for the global sample of ETGs is 0.32, so that the difference with the other slopes is significant.

As before the relation of non-homology can be defined only when an heterogeneous set of ETGs is used. For single classes of objects the relation is not well defined. In particular for BCGs and clusters, despite the values of $n$ are quite large ($2<n<8$ for BCGs and $0.4<n<5$ for clusters), the correlation is absent.

\begin{figure}
\includegraphics[scale=0.4,angle=0]{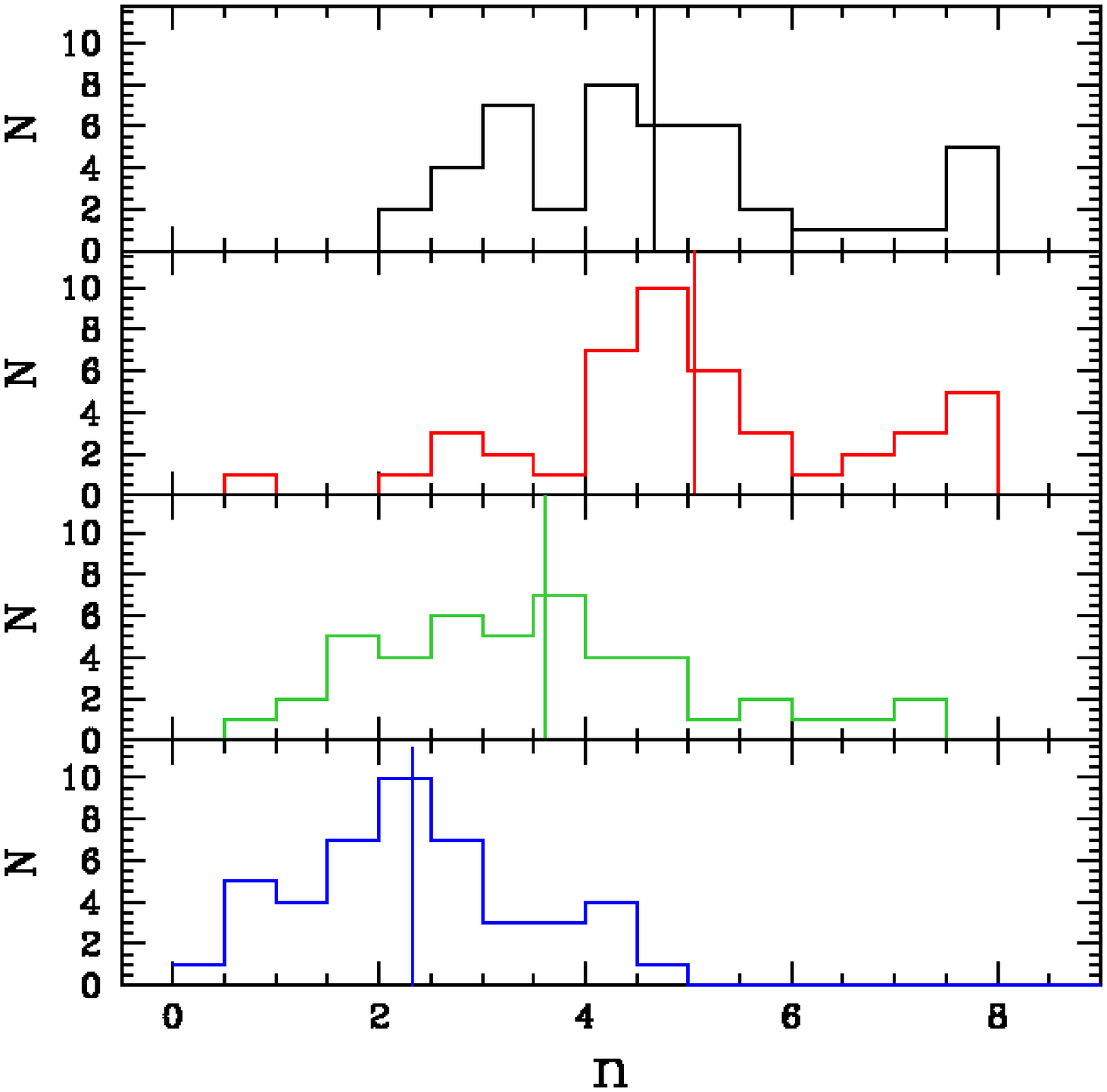}
	\caption{Histograms of the distributions of the S\'ersic index $n$ for the different classes of objects. The black histogram is for BCGs, the red for II-BCGs, the green for normal ETGs and the blue for clusters.
	The solid lines mark the average values of the distributions.}
	\label{fig:n}
\end{figure} 

Our clusters have approximately very similar luminosity, so that they do not show a clear relation between $n$ and \re\ or $L$. The different shapes of the light profiles however imply  that significant different structures are possible for clusters and bright galaxies of similar mass. In this respect we can consider clusters non-homologous systems as ETGs.

Fig.\,\ref{fig:n} shows the histograms of the distributions of the S\'ersic index $n$ for the different classes of objects. Note the spread of values of $n$ in each class and the increased value of $n$ for the brightest galaxies. 

In conclusion we can say that all our systems are non-homologous, in the sense that at any given mass might correspond quite different shapes.

This is the second clear parallelism.

\section{Comparison with numerical simulations} \label{sec:3}

Our aim here is to check to what extent current numerical simulations are able to reproduce the observed properties of real galaxies and clusters. We have chosen the data provided by the Illustris simulation\footnote{http://www.illustris-project.org/data/} \citep[][to whom we refer for all details]{Vogel2014,Genel_etal_2014,Nelsonetal2015}, a suite of large highly detailed hydrodynamics cosmological simulations, including star, galaxy and black-hole formation, tracking the expansion of the universe \citep{Hinshaw_etal_2013}.

We have used the run with full-physics, including both baryonic matter (BM) and DM having the highest degree of resolution, i.e. Illustris-1 \citep[see Table 1 of][]{Vogel2014}. A large number of different data are available for the subhalos inside this simulation. The former subhalos that we can call "simulated galaxies``, are those selected because the physics of baryons inside them can be studied in straight comparison with the observational data. Among the many tabulated quantities provided for galaxies, we work in particular with the \vfilt-band photometry, and with the mass and half-mass radii of stellar particles (i.e. integrated stellar populations), for which Cartesian comoving coordinates $(x',y',z')$ are available. 

We have analyzed the projected light and mass profiles using the $z'=0$ plane\footnote{The choice of the projecting plane does not alter the conclusion drawn below.} as reference plane and we have adopted the non-parametric morphology of \cite{Snyder_etal_2015}. Starting from the \vfilt\ magnitudes and positions of the stellar particles, we have computed the effective radius \re, the radial surface brightness profile in units of $r/R_e$, the best-fit S\'ersic index and the line-of-sight velocity dispersion. 

The values of \re\ are calculated considering only the star particles inside the Friend-of-Friends (FoFs) of galaxies and the galaxies inside the FoFs of clusters. We have set $z'=0$ to project the coordinates of stellar particles inside galaxies, so that the velocity dispersion is calculated along the $z'$ axis. 

Finally we exploit the Cartesian coordinates of DM particles to characterize the average distribution of the surface mass density: in this case we use the tabulated values of the half-mass radii to build up the radially normalized profiles of galaxies and clusters.  As for the real galaxies we have first normalized each profile to the effective radius and then calculated the average profiles of all classes of ETGs. The profiles have been built considering circular apertures in order to obtain the equivalent surface mass distribution.
For these profiles we have tested the S\'ersic's law as a fit of the projected mass distribution, providing the S\'ersic parameters of the best fits for the single galaxies and clusters. The same fits have been then applied to the average profiles.

To make easier the comparison with observations, we have built up the simulated data sets as follows: starting from the 20 most massive clusters (halo masses in the range from $2.77\times 10^{13}$ to $2.55\times 10^{14}\,h^{-1}\,M_{\odot}$) at redshift $z=0$, we have added three additional samples related to the galaxies inside clusters, namely 20 BCGs, 20 II-BCGs and a smaller set of normal ETGs. For these objects we have derived the same data available from observations.

The adopted procedure is the following. 
First, we get the data for the subhalos inside each cluster having non-zero stellar mass, that is equal to say having photometric data. At the end of this step, BCGs and II-BCGs can be trivially extracted as the first two luminous objects in each of the 20 catalogs. Note that the first and/or second most luminous objects may not be the first and/or second most massive object. 

For selecting the random ETGs we have used the non-parametric morphologies\footnote{These morphologies were derived for the $g$ band, but this does not alter the comparison with the \vfilt-band data used here.}. We use in particular two quantities: the Gini coefficient $G$, which measures the homogeneity of the light flux of a galaxy coming from different pixels, ranging from 0 (same flux from all pixels) to 1 (one pixel contains the flux of the whole galaxy) and $M_{20}$, which measures the second-order spatial moment of the pixels contributing 20$\%$ of the total luminosity \citep{Lotz_etal_04}. Starting from the 25 galaxies with available morphologies and with mass below $10^{10} M_{\odot}$ present in the simulation box at $z=0$ observed with Camera0 (i.e. the projection with $z'=0$ defined by \citealt{Torrey_etal_2015}), we have chosen the objects residing in the 20 most massive clusters with a morphology as close as possible to that of an ETG. This implies the following choice for the coefficients: $M_{20}<-2.00$ and $G>0.55$. With this criterion, we get a sample of only 5 ETGs.  This choice has been based on the visual inspection of Fig. 2 of \cite{Snyder_etal_2015} that clearly shows a well define limit between ETGs and Spirals at these values of the coefficients. Changing these limits we risk to bias our sample with a population of disk galaxies.

The details about these subhalos and the host cluster identifiers, together with the values of $M_{20}$, $G$ and the stellar masses are reported in Table \ref{tab:retg_vals} in Appendix. 

In Fig.\,\ref{fig:masses} we can see the histogram of the stellar mass of galaxy clusters inside the WINGS (solid black line) and Illustris (dashed blue line) sample (at $z=0$). The upper panel plots the mass inside $R_{200}$ while the lower panel that within $3R_{200}$. For the WINGS sample, the stellar mass has been computed for 31 over 46 clusters, i.e. for those having a proper estimation of the mass and a S\'ersic model with index $0.5<n<8$. We get the stellar mass using a constant $M^*/L$ ratio in the outermost regions of the clusters \citep[see][]{Cariddietal2018}.  This ratio can be followed along the whole cluster profile because the stellar mass of the more massive galaxies is known from the analysis of their spectral energy distribution \citep{Fritzetal2007,Fritzetal2011}.

For the Illustris sample, we summed up all stellar mass particles inside $r=3R_{200}$. As it can be seen from the figure, observed clusters are more massive (at least by a factor of 10) than the simulated ones, both at $r=R_{200}$ and $r=3R_{200}$, with the distribution of the simulated data being narrower than for real clusters.
Moreover, stellar masses of simulated clusters do not increase significantly while varying the enclosing radius: this may be due either to a higher mass concentration of simulated clusters with respect to the real ones, or to an 
over-efficiency of the feedback effects on the star formation of galaxies
\citep[for a detailed overview of drawbacks in Illustris related to feedback, see e.g. ][]{Vogel2014,Genel_etal_2014,Sparre_etal_2015}.
This is the first discrepancy we observe between real and artificial data.

\begin{figure}
\includegraphics[scale=0.4,angle=0]{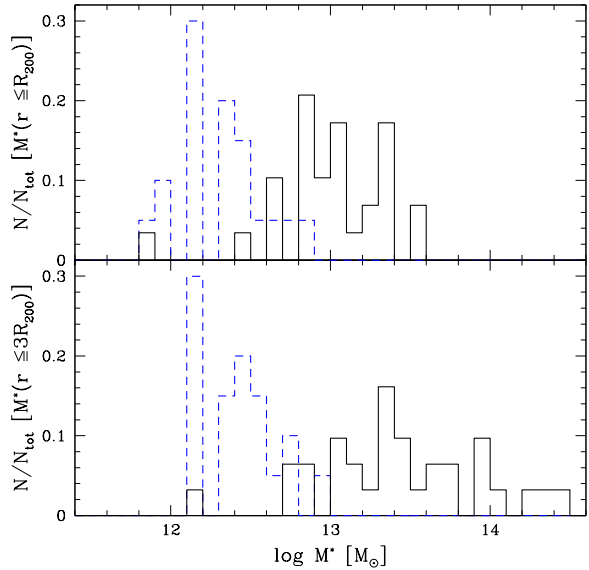}
	\caption{Histograms of the stellar mass distribution of real (black line) and simulated clusters (blue dashed line). The upper panel shows the comparison of the mass enclosed within $r=R_{200}$, while the bottom panel that inside $r=3R_{200}$.}
	\label{fig:masses}
\end{figure}

\begin{figure}
\includegraphics[scale=0.45,angle=0]{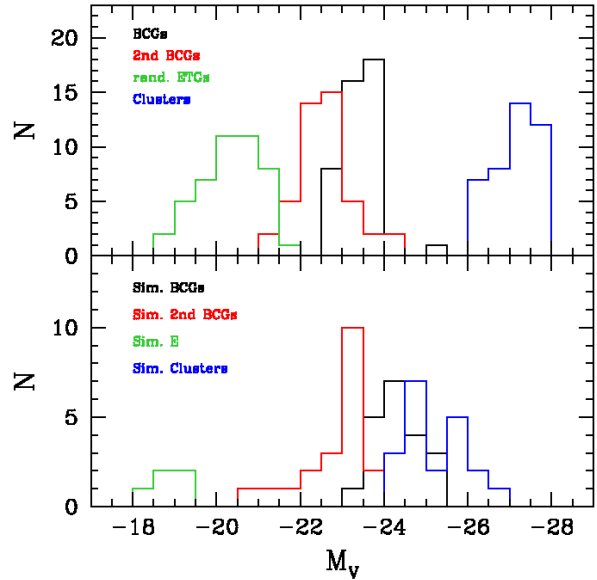}
	\caption{Histograms of the absolute \vfilt-magnitudes of real (upper panel) and simulated (bottom panel) objects.}
	\label{fig:magnitudes}
\end{figure}

Fig.\,\ref{fig:magnitudes} shows the absolute \vfilt-magnitudes of real (upper panel) and simulated (bottom panel) objects. Note again the large discrepancy between simulated and real clusters. Simulated BCGs appear a bit systematically brighter than real BCGs. The II-BCGs are a bit brighter too, while for normal ETGs the statistics is poor (but note that the five objects we get are fainter than normal real ETGs).

\begin{figure}
\includegraphics[scale=0.4,angle=0]{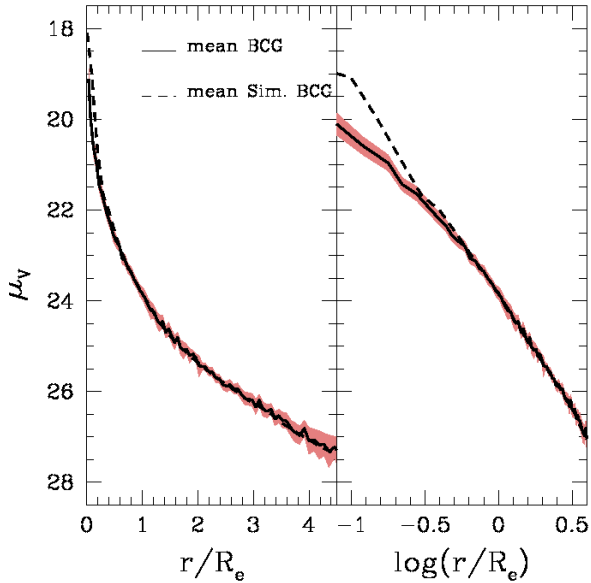}
	\caption{Comparison of the mean equivalent surface brightness profile of simulated BCGs (dotted line) with respect to real BCGs (solid line). The simulated profile has been artificially shifted in surface brightness to match the observed data. Left panel: $r/R_e$ in linear scale. Right panel: $r/R_e$ in log scale. The color band marks the 1$\sigma$ standard deviation around the mean BCG profile.}
	\label{fig:BCG-Sim}
\end{figure}

In Fig.\,\ref{fig:BCG-Sim} we see the comparison between the average profile of our BCGs and the equivalent average profile derived for simulated BCGs. The solid (dashed) line marks the observed (simulated) data. Once shifted in surface brightness for a constant value the agreement is excellent with the only exception of the inner region inside 0.5\re.  Only a small part of this difference might be attributed to seeing effect. One should in fact consider that the effective radius of BCGs is several times larger than the PSF. For the WINGS clusters the FWHM of the seeing was around 1 arcsec, that at the typical distance of the clusters corresponds to about 1 kpc (a radius from 10 to $\sim100$ times smaller than the typical \re).
This means that when we scale the profile in units of $r/R_e$ the region of influence of the PSF is always much lower than $0.1 r/R_e$. The difference we see in the right panel of Fig.\,\ref{fig:BCG-Sim} starts at $r/R_e=0.5$ and reach $\sim1$ mag at $r/R_e=0.1$. We therefore conclude that it could not be attributed to seeing. More likely simulated data are still not very precise in the region of the SMBH \citep[see e.g.][for a discussion in terms of stellar masses and the stellar mass function of galaxies]{Vogel2014}.

\begin{figure}
    \centering
\includegraphics[scale=0.4,angle=0]{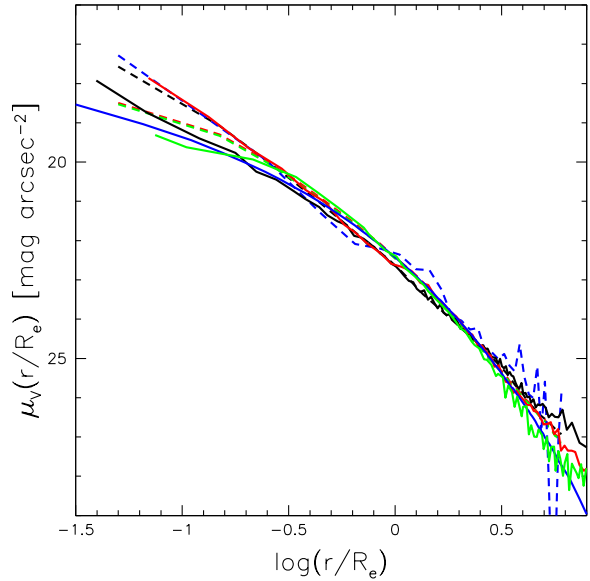}
    \caption{Average surface brightness profiles for observed (solid lines) and simulated structures (dashed lines) at $z=0$. BCGs are in black color, II-BCGs in red, normal ETGs in green and clusters in blue. The profiles have been vertically shifted by a constant value with respect to the BCG profile used as reference. See text.}
    \label{fig:avesc}
\end{figure}

In Fig.\,\ref{fig:avesc} we present the average surface brightness profiles for all our structures at $z=0$ in log scale.
The solid lines are used for the observed objects keeping the same colors of previous figures for BCGs, II-BCGs, normal ETGs and clusters. The dashed lines are used for simulated galaxies and clusters of the Illustris-1 dataset. 
Each profile has been superposed to the simulated BCGs by imposing the vertical shift calculated by eq. \ref{eqshift}.
The values of the shifts are listed in Table \ref{tab:shift_simbcg}.

\begin{table}[]
    \centering
    \caption{Values of the constant shift $s_j$ used to superpose the average surface brightness profiles of
    Fig.\,\ref{fig:avesc}. The average profile used as reference is that of our simulated BCGs.}
    \renewcommand{\arraystretch}{0.8}
    \begin{tabular}{|c|c|}
    \hline
    Obs. obj. & $s_j$ \\
    \hline
    BCG    & -1.2 \\
    II-BCG & -0.4 \\
    ETG   & -1.3 \\
    CL     & -7.8 \\
    \hline
    Sim. obj. & $s_j$ \\
    \hline
    II-BCG & 1.9 \\
    ETG   & 0.9 \\
    CL     & -3.3 \\
    \hline

    \end{tabular}
    \label{tab:shift_simbcg}
\end{table}

%%%%%

Again, the superposition of the profiles at $r/R_e \in [0.3,3.0]$ is remarkable, especially if we consider that the physical values of \re\ span more than two decades. The log scale permits to better see the differences in the inner region.
We observe that the II-BCGs depart from BCGs in both simulations and observations although in two
different fashions: the observed II-BCGs are on average more cuspy than BCGs, at variance with simulations in which the opposite is true. 
In the left panel of Fig.\,\ref{fig:Innprof} we have shown the $1\sigma$ uncertainty around the mean profiles for the various classes of objects, so when we compare these two figures, we can say that the difference between real and simulated BCGs and II-BCGs is only marginally significant.
However, the opposite trend seen in the inner regions in observations and simulations should be in some way related to the recipe adopted to follow the growth of the SMBHs and the AGN feedback.

Looking at random ETGs (green lines) we can see that the simulated profile stands very close to real galaxies and clusters, although the simulated average profile is slightly more cored in the center.

Clusters have instead significant differences. The average simulated cluster profile is much steeper in the center than that of real clusters and in the outer parts the profile is very noisy and seems a bit smoother than observed. Clearly, as noted above, simulations do not reproduce sufficiently well the cluster properties.

The non perfect match of the light profiles of real and simulated objects is the second discrepancy between observations and simulations. 

We now apply the Anderson-Darling criterion developed by \cite{Scholz&Stephens1987} to test the hypothesis that the observed distributions in the 1D surface brightness light profiles $\mu_V(r)$ have a common origin. We make the two-sample test rather than the three-sample test that just gives an overall value. 
For the comparison between BCGs and normal ETGs we get a p-value of 0.006, meaning a 0.6\% chance that the two samples come from the same distribution. We can then reject the common distribution hypothesis safely.
For the II-BCGs versus the normal ETGs we get a p-value of 0.009, meaning a 0.9\% chance so that we can also reject the common distribution hypothesis.
For BCGs and II-BCGs the p-value is 0.94, meaning a 94\% chance that the two samples come from the same distribution. The common distribution hypothesis is thus supported. 

The classical Kolmogorov-Smirnov test, that is differently sensitive in the tails of the distribution, gives a p-value of 0.003 for the comparison BCGs-normal ETGs, 0.006 for the comparison II-BCGs-normal ETGs and 0.86 for the comparison BCGs-II-BCGs, confirming the substantial similarity of the mean profiles of BCGs and II-BCGs and discarding the hypothesis of a common origin of these objects with normal ETGs.

\subsection{Non-homology in simulations}
\label{SecSim}

\begin{figure*}
    \centering
	\gridline{\fig{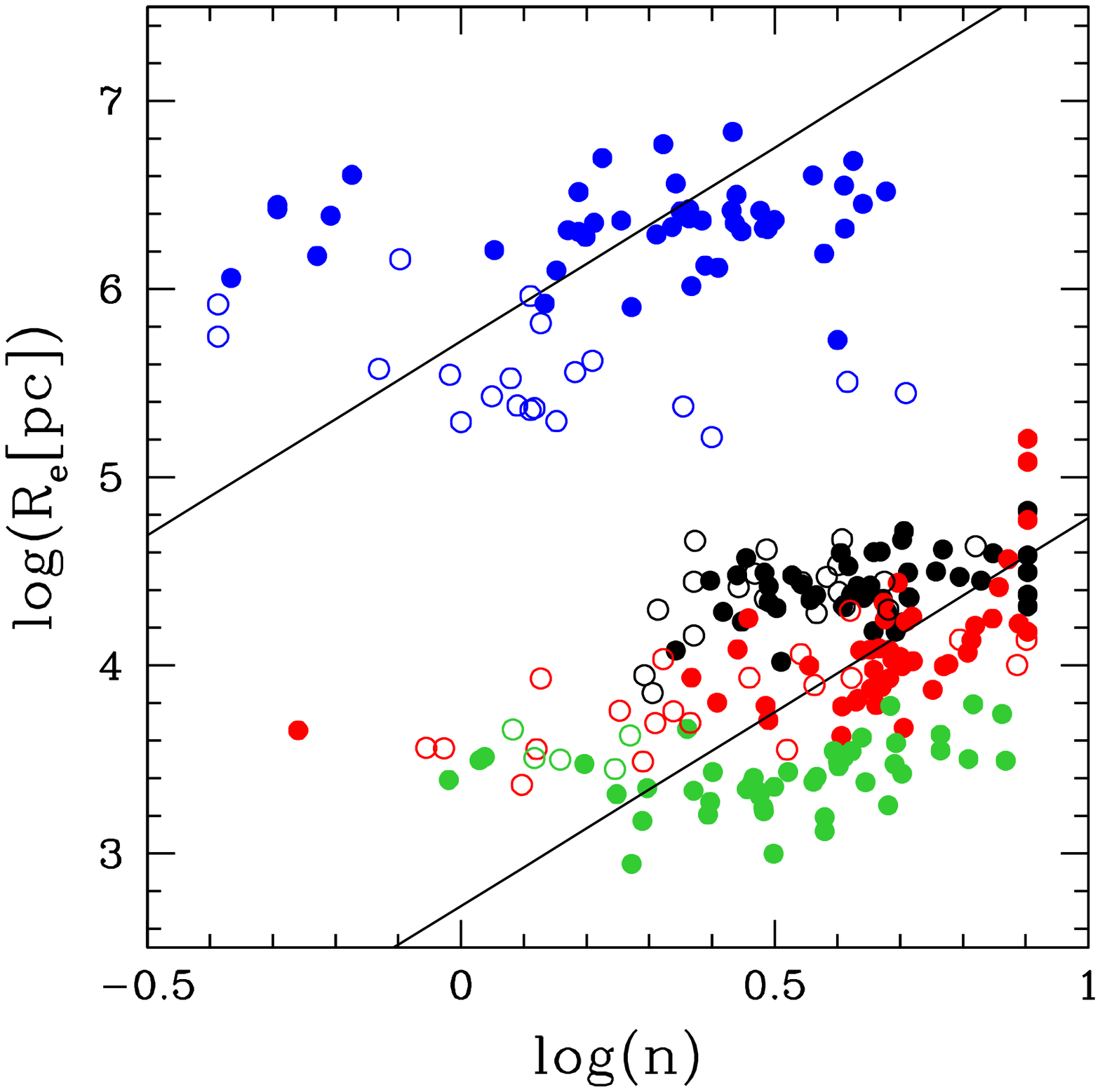}{0.5\textwidth}{}
		\fig{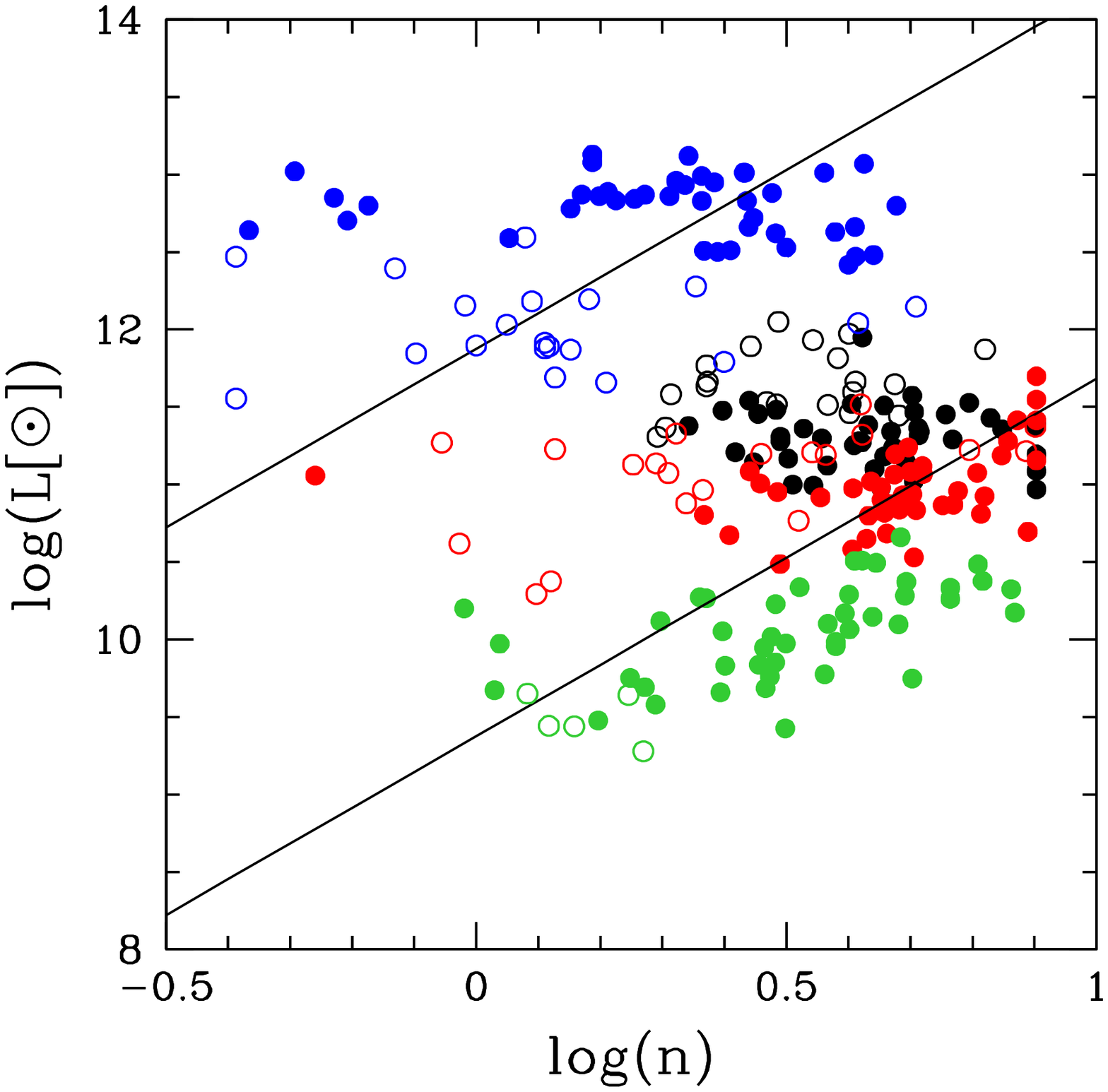}{0.5\textwidth}{}
	}
    \caption{Left panel: Distribution of ETGs and clusters in the \RnV\ plane. Black dots are BCGs, red dots are II-BCGs, green dots are random ETGs and blue dots are clusters. The simulated data have the same colors but are marked by open circles. The bottom black solid line marks the fit of the whole distribution of observed ETGs (BCGs+II-BCGs+random normal ETGs). The upper solid line is a shifted version of this fit used to best match the cluster distribution. Right panel: Distribution of ETGs and clusters in the \LnV\ plane. The symbols are the same of the left panel. The bottom black solid line marks the fit of the whole distribution of observed ETGs (BCGs+II-BCGs+random normal ETGS). The upper solid line is a shifted version of the bottom solid line done to best match the cluster distribution.
    }
    \label{fig:nRLs}
\end{figure*}

In order to check the non-homology we have fitted with the S\'ersic law the luminosity growth curves of the single profiles of BCGs, II-BCGs, normal galaxies and clusters extracted from simulations.  Figs.\,\ref{fig:nRLs} show real (filled symbols) and simulated (open symbols) objects in the same diagrams of Fig.\,\ref{fig:nR}. In both we see that BCGs and II-BCGs do show the same trend visible for real galaxies even if there is a systematic difference in luminosity and in radius (in particular for clusters).
The large interval in the values of $n$ confirms that simulations are able to reproduce the observed spread of shapes of galaxies and clusters and therefore that in general the structures emerging from merging events are not self-similar.

The values of $n$ measured for the averaged profiles of simulated galaxies are listed in Tab. \ref{Tabfitprofiles}. They are quite similar to that seen for real galaxies in the case of II-BCGs, but systematic variations are visible for the other classes. Taking into account how the final value of $n$ depends on several factors we can conclude that simulations successfully reproduce ETGs while fail with galaxy clusters.

\subsection{The progenitors of simulated galaxies}\label{Progenitors}

We have followed the SubLink merger trees of simulated galaxies \citep[see e.g. for details][]{Rodriguezgomez_etal_2015} to map the evolution of all the subhalos back in time. We use the relevant data at redshift $z=0$, $z=0.2$ (i.e. at a look-back time
$t_{lb}\sim2.6$ Gyr) and $z=0.8$ ($t_{lb}\sim 7.0$ Gyr). With these data we show here the evolution of the effective radius and of the half-mass radius of BCGs and II-BCGs up to redshift $z=0.8$.

\begin{figure}
\includegraphics[scale=0.45,angle=0]{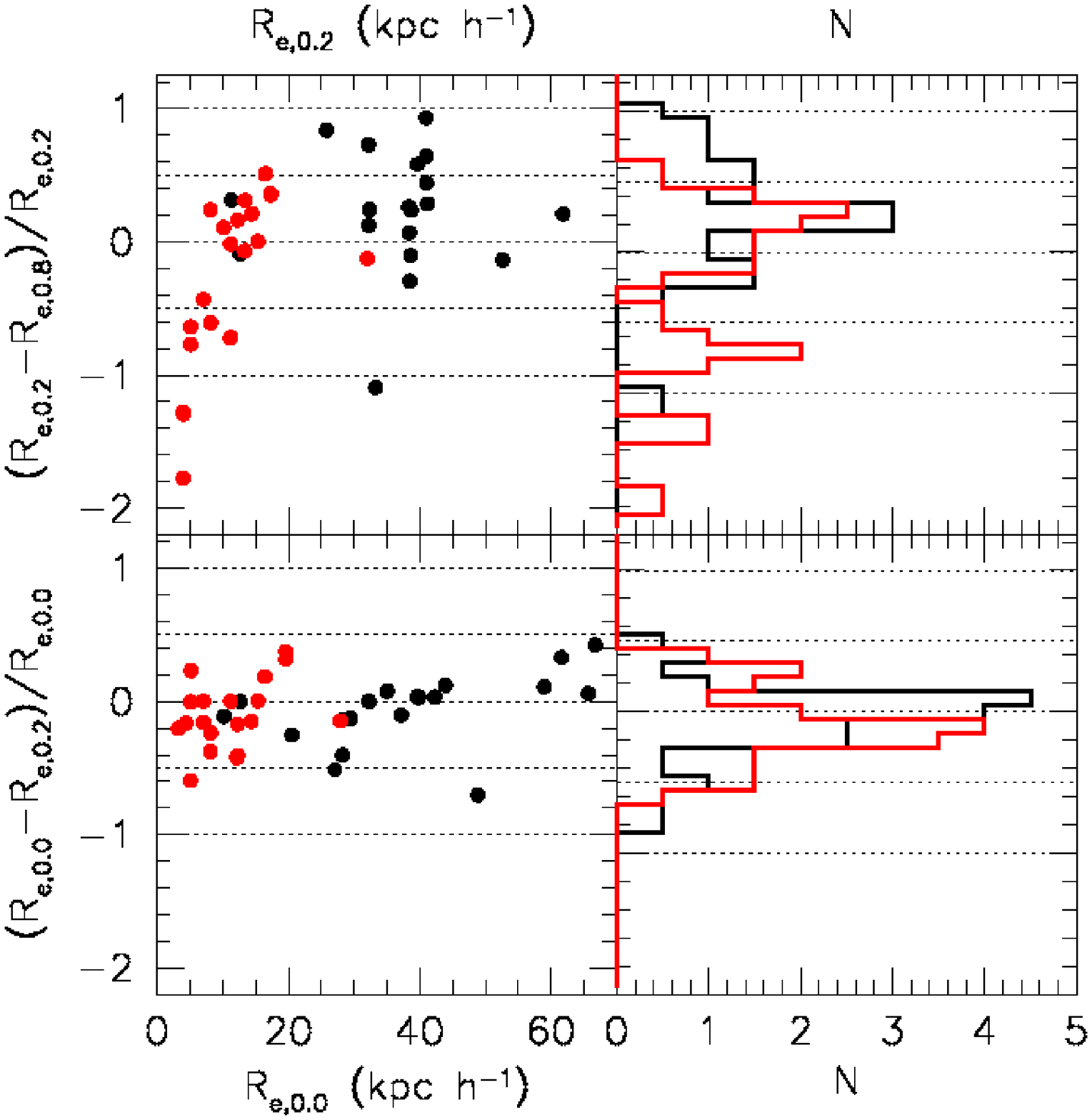}
	\caption{Evolution of $R_e$ for BCGs and II-BCGs in the Illustris simulation, evaluated at three reference epochs (upper panels: from $z=0.8$ to $z=0.2$; lower panels: from $z=0.2$ to $z=0.0$).
	In the left panels, we display the final values of \re\ against the differences (final-initial)/final; 
	black filled circles represent BCGs, red filled squares
	represent II-BCGs. In the right panels, the histograms show the overall behavior of the two datasets, with the black solid line representing BCGs and the red line II-BCGs. In both panels the dotted lines mark the relative variation of \re\ ($0\%$, $\pm 50\%$ and $\pm 100\%$).}
	\label{fig:sim_evo_re}
\end{figure}

The first thing we want to stress is that the progenitors of present BCGs are also BCGs at that epochs \citep[BCGs are known to be already in place at $z\sim 1$; see e.g.][]{DeLuciaBlaizot2007}, while the progenitors of present II-BCGs are not necessarily the 2nd brightest object in the clusters at the past considered epochs. 
The II-BCG observed at $z=0$ is the final result of two competing effects: it is a galaxy that has experienced a fast growing through merging, surviving as II-BCG in the merger tree of the simulation, but at the same time is an object that lives in the innermost cluster region where disruptive phenomena are at work. This peculiar evolution of the II-BCGs will be much clear below when we will look at the evolution of the effective radius of galaxies.

In Fig.\,\ref{fig:sim_evo_re} we show the evolution of \re\ through three reference epochs for the 20 main progenitor branches of BCGs and II-BCGs. We display for each BCG (black filled circles and solid histograms) and II-BCG (red filled squares and dotted histograms) the variations of \re\ normalized to the final values reached at $z=0.2$ (upper panel), and $z=0.0$ (bottom panel).
The dotted lines mark the percentage of variation ($0\%$, $\pm 50\%$ and $\pm 100\%$).

We note that both positive and negative variations of \re\ are present either for BCGs and II-BCGs. 
This analysis aims at giving a profile-oriented insight on the average evolution of our objects.
In the top panel of Fig.\,\ref{fig:sim_evo_re}, BCGs and II-BCGs show two broad distributions, having their maximum peak where the radius changes by $\sim +30\%$.
Looking at BCGs, their effective radii tend most likely to increase,
with growths of $\gtrsim \%50$  in 5 cases over 20; the ample range of values of $\Delta R_e/R_e$ covered by BCGs passing from $z=0.8$ to $z=0.2$ sheds light on an important feature: although they can be considered as already assembled in terms of overall large mass and photometric output, structural properties like e.g. the concentration of stellar light (thus mass) are still evolving. 

II-BCGs have a more ample range of behaviors and also a number of minor peaks at negative values. In this context and with the chosen reference epochs, the high degree of complexity in the build-up of II-BCGs can be appreciated, being as already stated a particular subsample of galaxies evolving faster and more violently than others.
In the bottom panel, the histograms related to BCGs and II-BCGs are quite similar in terms of both width and value of the peak: this suggests that II-BCGs in the last $\sim 2.6$ Gyr have evolved and consolidated into a more homogeneous sample.

With the Kolmogorov-Smirnov test we have checked the hypothesis of evolution of the effective radius \re\ from $z=0.8$ to $z=0.2$ and $z=0$. For the BCG sample the comparison of \re\ between $z=0.8$ and $z=0.2$ we get a probability of 0.008, discarding the null hypothesis of an equal distribution of the radius at these epochs. On the other hand the comparison of \re\ between $z=0.2$ and $z=0$ gives a probability of 0.77, that implies an almost equal distribution of \re\ at these epochs. For the II-BCGs the test gives a probability of 0.27 in the comparison of \re\ between $z=0.8$ and $z=0.2$ and of 0.96 for the comparison of \re\ between $z=0.2$ and $z=0$.
For these objects the evolution of the effective radius is less marked, although again the more marked variation occurs between $z=0.8$ and $z=0.2$.

\begin{figure}
\includegraphics[scale=0.45,angle=0]{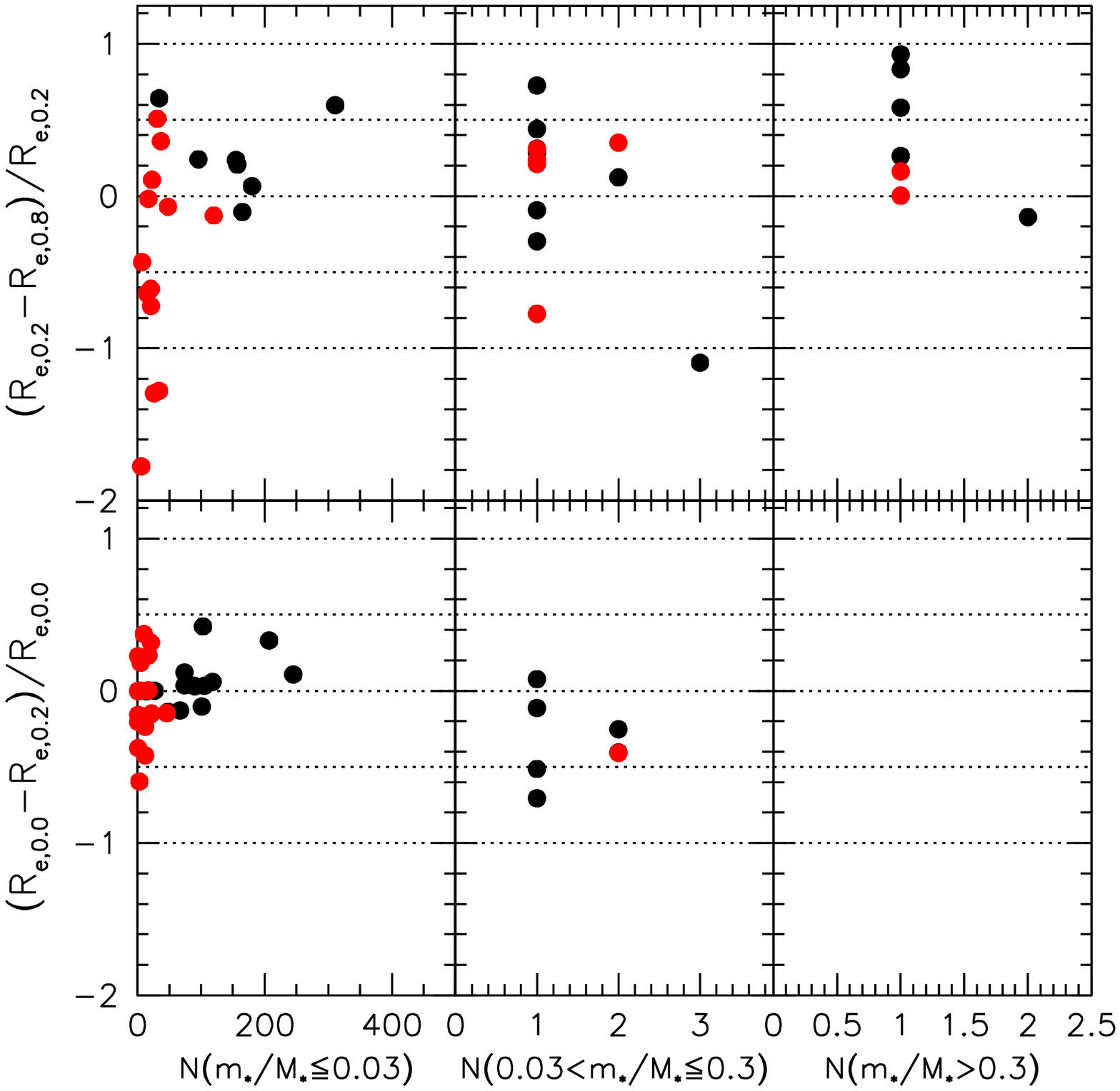}
	\caption{The relative variation of \re\ versus the number of merging events.  Black dots mark BGCs, while red dots II-BCGs. From left to right the mass ratio between donor and main progenitor increases.
	The dotted lines mark the percentage of variation ($0\%$, $\pm 50\%$ and $\pm 100\%$).}
	\label{fig:Nummerg}
\end{figure}

Finally Fig.\,\ref{fig:Nummerg} shows the same variations of \re\ as a function of the number of merging events that BGCs and II-BGCs have experienced. In the left panel we give the number of small merging events, expressed as the ratio of the mass $m^*$ of the donor with respect to that $M^*$ of the main progenitor. Going to the right panels we observe that the number of big merging events is null today.

In conclusion we can say that both positive and negative variations of \re\ are observed in the hierarchical simulations, but a net average variation of \re\ is present from $z=0.8$ and $z=0.2$. At closer epochs the average radius of BCGs and II-BCGs does not change very much. 

\subsection{Surface mass density profiles} \label{sec:surfacemass}
As already claimed by \citet{Merrittetal2005} the S\'ersic's law can give an equally optimum fit of the surface mass density profile as the NFW. 
In that paper they noted that the energy distribution of the S\'ersic's law is roughly Boltzmann and they argue that this is a “maximum-entropy” state resulting from mixing occurred during violent relaxation or merger events \citep[see also][]{Binney1982,MerrittTremaineJohnstone1989,Ciotti1991}. 
Since the DM halos are also well mixed and approximated by a power-law \citep{TaylorNavarro2001}, they
suggested that the scale-free property of the S\'ersic's law is a feature in common between dark and luminous spheroids. In other words the S\'ersic's law is a "universal profile" for all types of structures formed by merging.

In the following we want to test whether the surface mass density profiles of our structures emerged from simulations are indeed well fitted by the S\'ersic's law.
We have therefore repeated the analysis done for the surface brightness profiles looking now at the DM and DM+StM (StM=stellar matter) distributions inside BCGs, II-BCGs and clusters.

For the sake of clarity in this work we use the following definition of surface mass density:

\begin{equation}
\mu_m(r) \equiv -2.5\log \left(\frac{1}{2\pi r}\frac{dM}{dr}\right).
\end{equation}

As for the surface brightness, this quantity grows towards positive values in presence of less dense 2D regions. We evaluated $\mu_m(r)$ inside circular annuli of radius between $r$ and $r+dr$ and, as we did for the photometric data, projecting the mass on the $z=0$ plane.

\begin{figure}
\includegraphics[scale=0.4,angle=0]{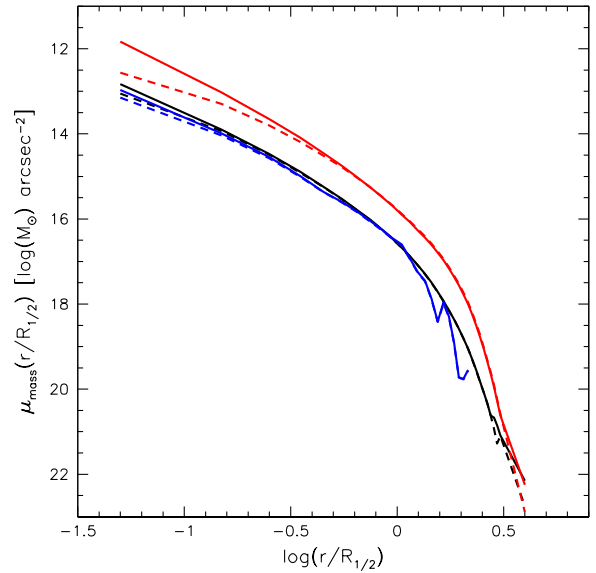}
	\caption{The average surface mass profiles $\mu_m$ for BCGs (black lines), II-BCGs (red lines) and clusters (blue lines) at
	$z=0.0$ from simulations; dashed lines refer to average profiles of DM only, while solid lines refer to DM plus stellar
	particles. No rescaling has been made in this plot.}
	\label{fig:sim_avecz00}
\end{figure}

\begin{figure*}
    \centering
	\gridline{\fig{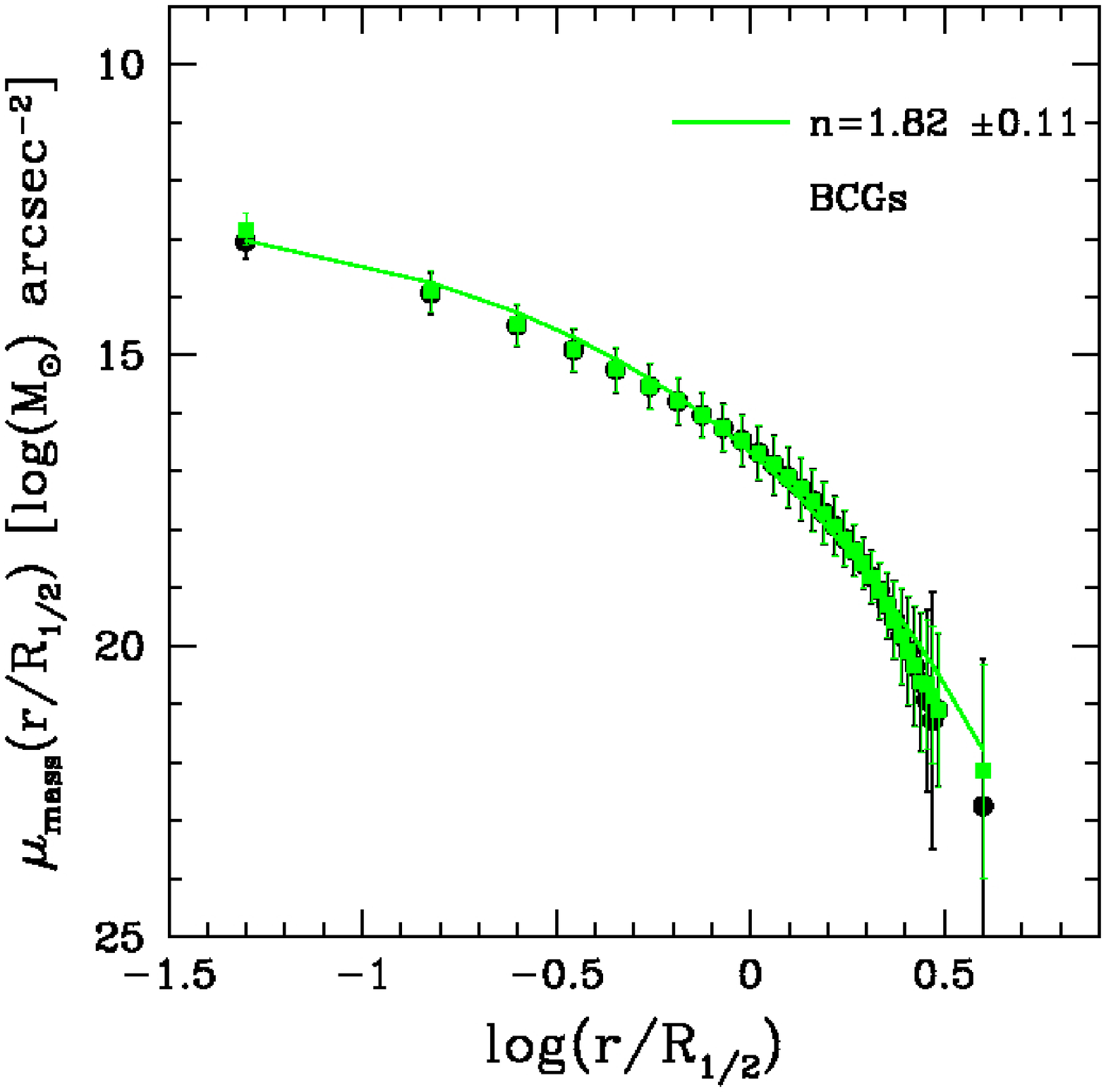}{0.32\textwidth}{}
		\fig{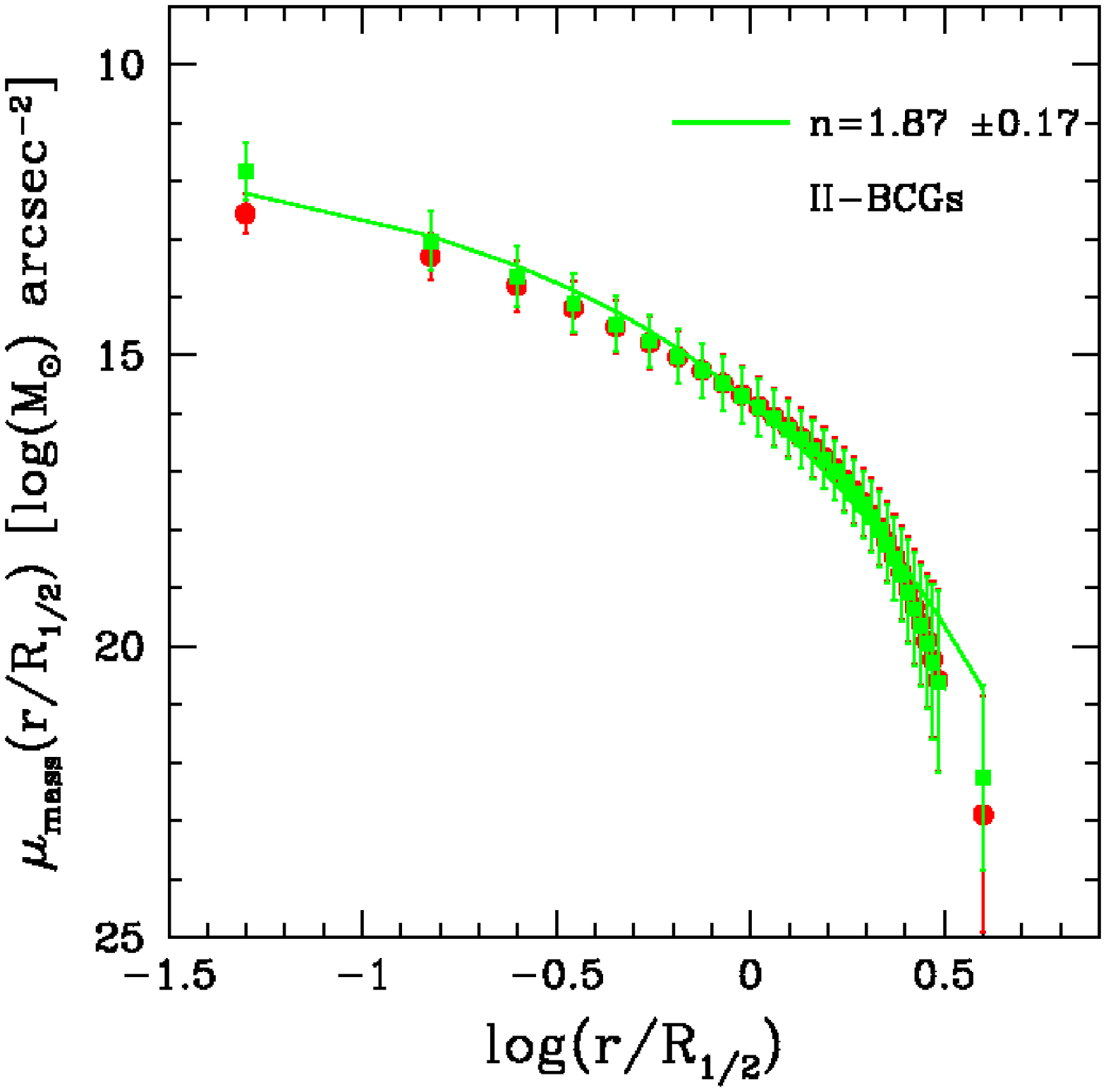}{0.32\textwidth}{}
		\fig{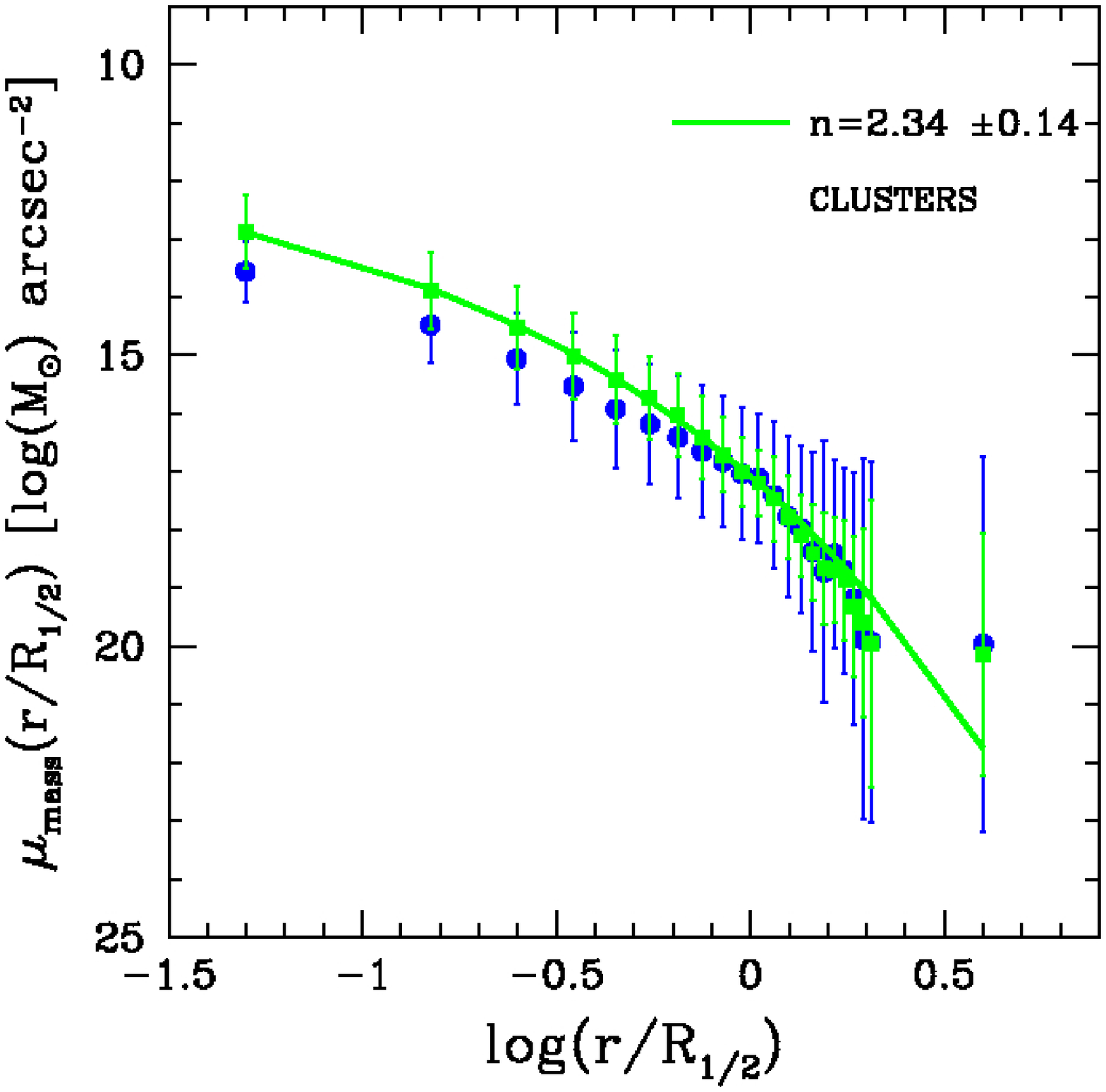}{0.32\textwidth}{}
	}
    \caption{The S\'ersic fit for the average mass profiles of BCGs (left panel), II-BCGs (central panel) and clusters (right panel).
    Error bars mark the $1\sigma$ rms of each radial bin. Green dots are used for the DM+StM matter while black, red and blue dots are used for the DM in BCGs, II-BCGs and clusters respectively. The S\'ersic index listed in the box refer to the fit of the DM+StM component shown by a green line.}
    \label{fig:sim_sers_mass}
\end{figure*}

We begin by presenting the averaged radial profiles at $z=0$ in Fig.\,\ref{fig:sim_avecz00}. The dashed lines refer to DM-only profiles, while the solid lines are those of DM+StM.  Black lines are used for BCGs, red lines for II-BCGs and blue lines for clusters.
$R_{1/2}$ is determined by considering not only the matter distribution, but also the relative abundance of each species. The main contributor is therefore DM, stellar particles and other species (e.g. gas, black holes) contributing only marginally. For these reasons, $R_{1/2}>R_e$ for each structure, and consequently the observed distributions have a much more important drop-off in density at relatively lower values of the normalized radial distance. 

For clusters, in order to keep the BM into
account we have included the galaxies embedded into the DM subhalos. 

At variance with Fig.\,\ref{fig:avesc}, no rescaling has been made in plotting the average profiles.
One trivial consideration looking at Fig.\,\ref{fig:sim_avecz00} is that the inclusion of stellar particles leads to steeper profiles especially in the central regions, because more matter is included in the same radial bins. In general the shapes of the profiles are quite similar, with II-BCGs standing above BCGs and clusters (see also Fig.\,\ref{fig:Allprof2}). 

The first thing to stress is that now the projected mass profiles are normalized to the half-mass radius $R_{1/2}$, which is related to all kinds of massive particles in the simulation.

The second important thing to note is that, at variance with light profiles, the surface mass density profiles begin to be very noisy for $r>3R_{1/2}$. This is also true for the average profiles. The reason is that the lower is $R_{1/2}$ the less is the mass present in the outskirts, and in turn the greater the impact of mergers in shaping the outermost region of a profile. 
For this reason our analysis is focused on the region $r\leq 3R_{1/2}$.

Fig.\,\ref{fig:sim_sers_mass} shows the results of the fits obtained for the surface mass density of BCGs (left panel), II-BCGs (central panel) and clusters (right panel). The average mass density is evaluated at $z=0$.

We have separated the fits of DM+StM profiles (green squares), to that of pure DM profiles (black, red and blue circles for BCGs, II-BCGs and clusters respectively). For clusters 
we plot the average profile in the range $[0,\,2R_{1/2}]$. 

An important thing to keep in mind is that adding galaxies inside clusters is not the same of adding stellar mass inside galaxies: the distribution of galaxies inside clusters could be very different in the outermost regions. It follows that often the
DM+StM profiles can be quite different from the DM-only profiles in particular in the outer radial bins.

This source of noise in clusters motivates us to fit the average profile of clusters only for $r<R_{1/2}$. 

Fig.\,\ref{fig:sim_sers_mass} indicates that the projected mass distributions follow the S\'ersic's law quite well:
the fits are well within the error bars. 
In Tab. \ref{tab:sersicmass_rmscomp} we list the values of the best fits obtained for the various classes of objects at different cosmic epochs.

\begin{table*}[]
    \centering
    \renewcommand{\arraystretch}{0.8}
    \caption{Best S\'ersic parameters for the average surface mass density profiles of simulated BCGs, II-BCGs and clusters at the three selected cosmic epochs in the ranges $r/R_{1/2} \in [0,1]$ (subscript 1) and $r/R_{1/2} \in [0,3]$ (subscript 3).  The first three lines provide the fits for BCGs at $z=0$, $z=0.2$ and $z=0.8$ respectively. The second three lines give the same for II-BCGs. Clusters are fitted only at the present epoch. The errors $\Delta n$ and $\Delta b$ related to the parameters are given as well as the values of the $1\sigma$ scatter obtained in the fits of the single and averaged profiles.}
    \begin{tabular}{|c|c c c c | c c c c | c | c|}
    \hline
         & $n_1$ & $\Delta n_1$ & $b_1(n)$ & $\Delta b_1(n)$ & $n_3$ & $\Delta n_3$ & $b_3(n)$ & $\Delta b_3(n)$ & $\sigma_{m(1,3)}$ & $<\sigma_{m(1,3)}>$ \\
    \hline
    BCGs@z=0 & 3.87 & 0.17 & 6.30 & 0.18 & 1.82 & 0.11 & 4.15 & 0.22 & 0.01, 0.11 & 0.28, 0.25  \\
    BCGs@z=0.2 & 3.76 & 0.21 & 6.10 & 0.21 & 1.80 & 0.10 & 4.10 & 0.20 & 0.01, 0.11 & 0.24, 0.23 \\
    BCGs@z=0.8 & 3.86 & 0.24 & 6.10 & 0.25 & 1.33 & 0.08 & 3.28 & 0.22 & 0.02, 0.12 & 0.35, 0.31 \\
    II-BCGs@z=0 & 4.68 & 0.26 & 7.64 & 0.31 & 1.87 & 0.17 & 4.14 & 0.35 & 0.02, 0.24 & 0.34, 0.28 \\
    II-BCGs@z=0.2 & 6.28 & 0.16 & 9.52 & 0.19 & 1.70 & 0.16 & 3.94 & 0.34 & 0.02, 0.21 & 0.37, 0.29 \\
    II-BCGs@z=0.8 & 6.84 & 0.20 & 10.01 & 0.24 & 1.83 & 0.18 & 3.91 & 0.34 & 0.01, 0.22 & 0.33, 0.26 \\
    CLs@z=0 & 2.35 & 0.14 & 5.35 & 0.17 & 4.11 & 0.95 & 7.85 & 1.78 & 0.04, 0.45 & 0.52, 0.70 \\
    \hline
    \end{tabular}
    \label{tab:sersicmass_rmscomp}
\end{table*}

Note that the values of $n$ are quite different according to the selected intervals in radius for the fits. With the exception of clusters the values of $n$ are lower than 2 when $r/R_{1/2} \in [0,3]$.
We do not see anymore the progressive decrease of $n$ visible in the fits of the luminous component in BCGs, II-BCGs and clusters.

The rms deviation between the average profiles and their S\'ersic fit was calculated as follow:

\begin{equation}
\begin{aligned}
\sigma_m^2 &\equiv <\Delta\mu_m^2> \\
           &= \frac{1}{N}\sum_{i=1}^{N} \left(\mu_m(r_i/R_{1/2})-\mu_{m,fit}(r_i/R_{1/2})\right)^2\,.\\
\end{aligned}
\label{eqrms}
\end{equation}

A direct comparison with the $\sigma_m$ found by
\citet{Merrittetal2005} is not easy because they fitted the S\'ersic's law on the DM-only mass profiles using a normalized radial coordinate $X=R/R_e$ [see their Equation (1)]. In this work, the S\'ersic fit is done for the DM+StM mass 
profiles and the radial coordinate is normalized to half-mass radius $R_{1/2}$. 

As already seen for the surface brightness profiles, the radial extension of the S\'ersic fit and the normalization parameter are crucial for determining the final value of $n$.

In Table \ref{tab:sersicmass_rmscomp} we also report the rms evaluated by eq. \ref{eqrms}. The last two columns list the values of $\sigma_m$ obtained by fitting the individual profiles in the range $r/R_{1/2} \in [0,1]$ (subscript 1) and $r/R_{1/2} \in [0,3]$ (subscript 3) and the average profiles ($<>$) in the same ranges. 
Galaxies at all reference epochs show the minimum rms when fitted in the range $r/R_{1/2} \in [0,3]$, while clusters are best fitted only in the range $r/R_{1/2} \in [0,1]$, because at larger radii the uncertainties are too large.

In the Appendix we provide from Table \ref{tab_sersic_single_mass_1} to Table \ref{tab_sersic_single_mass_7} 
the least-squares fit values of the single profiles
for galaxies and clusters. From these Tables we derive the mean values of $\sigma_m$ reported in the last column of Table \ref{tab:sersicmass_rmscomp}. 
Note that the rms values of the single profiles are higher than those related to the average profiles
and that the errors related to the S\'ersic 
parameters are lower than 10$\%$. 

\begin{figure*}
    \centering
	\gridline{\fig{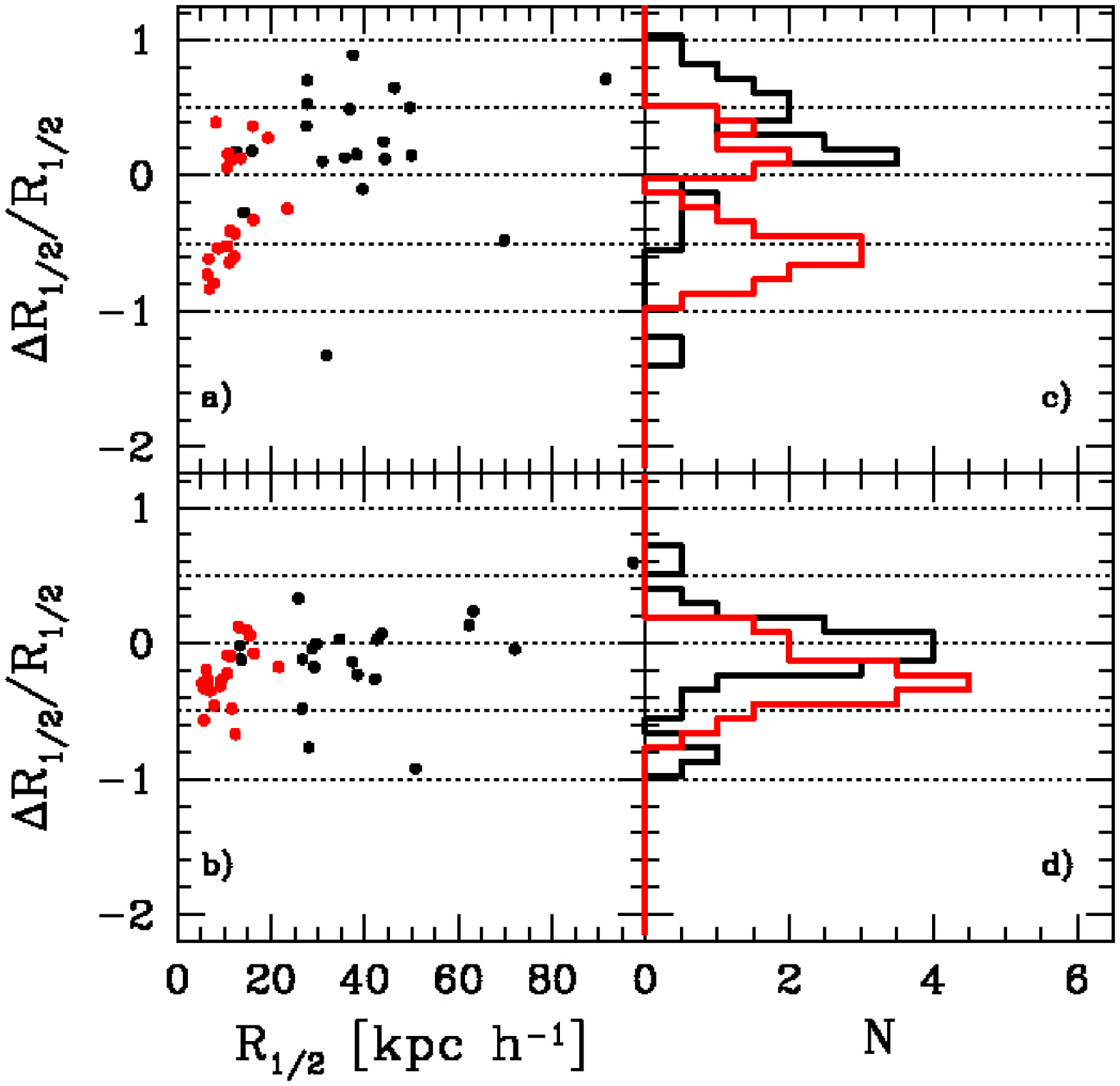}{0.42\textwidth}{}
		\fig{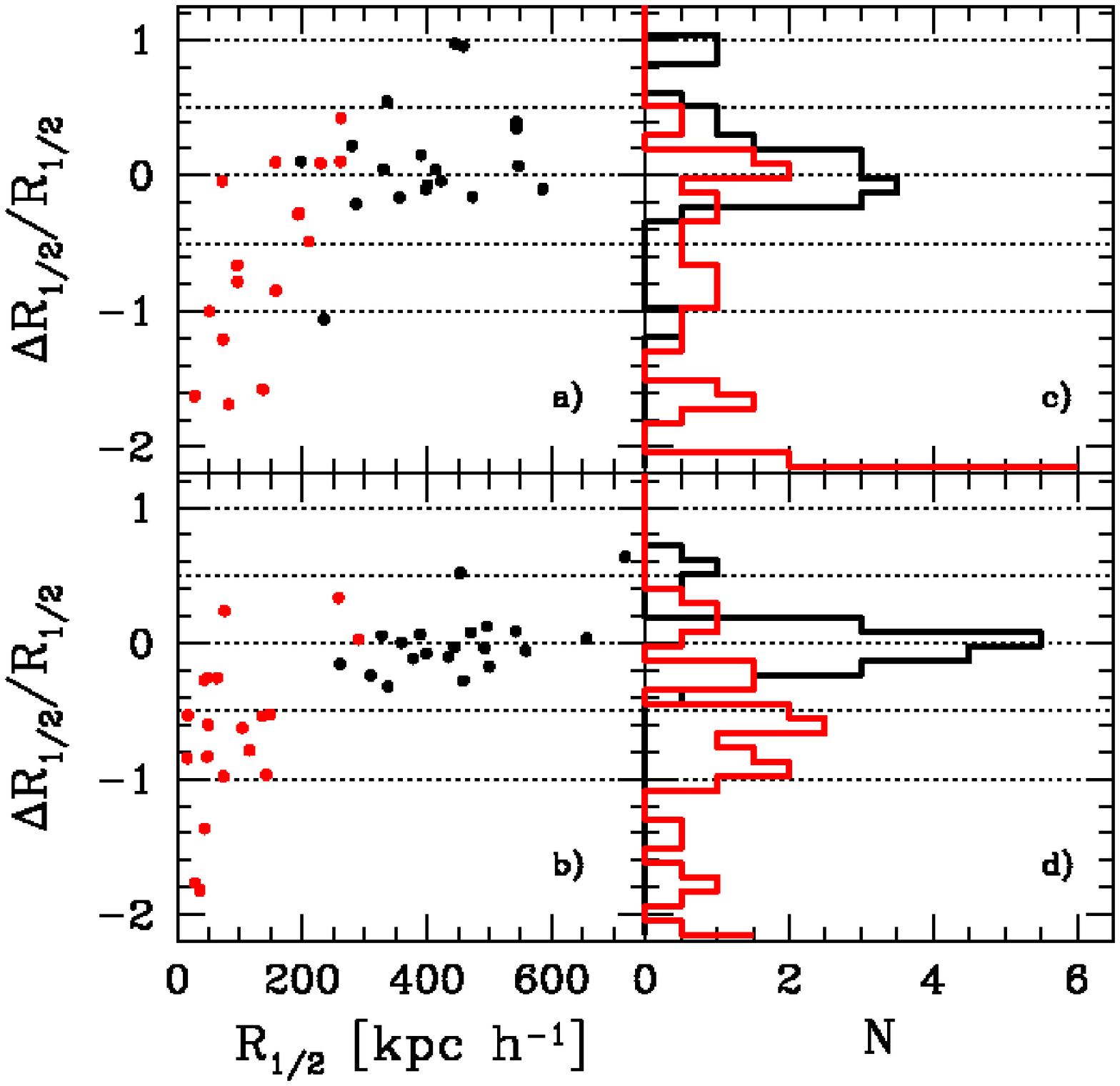}{0.42\textwidth}{}
	}
    \caption{Left panel: In panel a) the evolution of the half-mass radius considering only the contribution of the StM. The panel shows the relative variations of the half-mass radius evaluated at $z=0.8$ and $z=0.2$. Black dots refer to BCGs and red dots to II-BCGs. In Panel b) the same variation for the redshift interval $z=0.2$ to $z=0$. In Panels c) and d) the histograms of the relative half-mass variation at the selected epochs. The black line is used for BCGs and the red line for II-BCGs. Right panel: the same plot of the left panel, but using the half-mass radii obtained when the DM+StM matter inside galaxies is considered.}
    \label{fig:sim_evo_hmrst}
\end{figure*}

In order to see how the normalization radius evolve in the selected epochs, we replicate  Fig.\,\ref{fig:sim_evo_re} in Fig.\,\ref{fig:sim_evo_hmrst} using the half-mass radius instead of \re. The relative variations of the half-mass radius, either considering the StM (left panel) or the DM+StM (right panel), are quite similar. Both positive and negative variations are visible as in the case of \re.

Again with the Kolmogorov-Smirnov test we have checked the hypothesis of evolution of the half-mass radius $R_{1/2}$ from $z=0.8$ to $z=0.2$ and $z=0$. For the BCG sample considering only the contribution of the StM the comparison of $R_{1/2}$ between $z=0.8$ and $z=0.2$ we get a probability of 0.02, discarding the null hypothesis of an equal distribution of the radius at these epochs. On the other hand the comparison of $R_{1/2}$ between $z=0.2$ and $z=0$ gives a probability of 0.96, that implies an almost equal distribution of $R_{1/2}$ at these epochs. When we consider the DM+StM the values of the probability are 0.27 and 0.50 respectively for the two redshift intervals, indicating a mild evolution of the half-mass radius.
For the II-BCGs with StM only the test gives a probability of 0.27 in the comparison of $R_{1/2}$ between $z=0.8$ and $z=0.2$ and of 0.77 for the comparison of $R_{1/2}$ between $z=0.2$ and $z=0$. With DM+StM we get instead 0.06 for both time intervals. In this last case the test seems to indicate a larger probability of variation of the half-mass radius.

When we look at the distribution of the total mass
(right panel of Fig.\,\ref{fig:sim_evo_hmrst}) we remember that the
DM is now the chief contributor; still very dispersed histograms are visible for the II-BCGs both in the top and bottom panels. This clearly indicates that the assembly in terms of overall mass is still occurring at the selected epochs. In other words the II-BCGs are much younger objects than BCGs, whose radius is almost no evolving. 

In conclusion we can say that the S\'ersic's law is valid both for the luminous and dark components of all type of structures (both galaxies and clusters). The values of $n$ however depend very much on the choice done for the fitting interval and on the normalization radius used. The values of $n$ are quite similar for the mass profiles, in particular when the large interval in radius is considered, while are different for the luminous components according to the stellar mass of the system. 

The idea of a universal profile implies that all structures start from subhalos of similar shapes and then evolve differently for the activity of the baryon component. This last depends not only on the stars formed, but also on the merging history and the feedback effects that might be quite different from galaxy to galaxy.  

\section{Conclusions and discussion} \label{sec:4}

This paper is the first of a sequel dedicated to the analysis of the parallelism of properties shown by clusters and ETGs. In particular we have analyzed the behavior of their luminosity and mass profiles. 
By exploiting the data of the WINGS and Omega-WINGS surveys we have created the average equivalent luminosity profiles for different classes of objects: BCGs, II-BCGs, normal ETGs and clusters and we have checked the degree of non-homology of such systems. Then, using the data of the "Illustris" simulation we have tested the ability of current hydrodynamical models in reproducing the observational data. These data have also permitted the analysis of the surface mass profiles and a check of the effective radius of ETGs at higher redshifts.

In summary these are our main conclusions:

\begin{itemize}
\item The equivalent observed luminosity profiles in the \vfilt-band of all classes of objects, once normalized to the effective radius and shifted in surface brightness, can be superposed with a small scatter ($\le0.3$ mag). 

\item The average profile of each class slightly deviates from the other in particular in the inner and outer regions. The Anderson-Darling test does not support the hypothesis that BCGs and II-BCGs come from the same population of normal ETGs, while supports the substantial similarity of the average profiles of BCG and II-BCGs. This means that, if ETGs start from a common universal profile, the subsequent evolution has almost completely changed the profile in the inner and outer regions, leaving unaltered the profile in the middle. We can then argue that all light profiles might originate from the same mass profile and later evolve for the numerous merging events and for the feedback effects of SNe and AGN. These events do not seem to have affected the bulk of the luminosity/mass distribution, but have likely had a key role in determining the shift of the profiles of the big galaxies with respect to that of normal ETGs. The presence of a SMBH in the center should have systematically changed the shape of the light profiles \citep{KormendyHo2013}. It is remarkable, although not fully statistically demonstrated, that the average II-BCGs profile is steeper than that of BCGs in the center, while the opposite is true in the outer regions. We can only speculate on the reason for this at the moment. We believe that this behavior should be connected with the number of merging events (wet and dry) that a system has experienced during its evolution. BCGs are at the center of the cluster potential so that they likely experienced several events of merging during their history. Some of them might have accreted SMBHs from other galaxies increasing the total mass in the center. The presence of large masses in orbits around each other might have contributed to eliminate many stars from the central area \citep{Bonfini2016}. This is the binary BH scouring scenario. 
Another possibility has been suggested by \cite{NipotiBinney} who argue that thermal evaporation of cold gas by virial-temperature gas might have played an important role in determining the shape of the inner profiles, since the analysis of the data reveal that galaxies with power-law profiles in the center have younger stellar populations than those with cores. 

In the outer regions, on the other hand, the big number of dry mergers might have produced the observed extended stellar halos of BCGs and II-BCGs, with the first systems being much more frequently affected by these events.

Notably galaxy clusters behave like normal ETGs and not as BCGs and II-BCGs. We argue that this is related to the merging history of clusters: as it occurs for low luminous ETGs, the number of dry merging events has not likely altered considerably the original light profile. 

\item We have shown that the S\'ersic's law provides good fits of the luminous and dark components of galaxies and clusters and that all systems are non-homologous, in the sense that their shapes cannot be predicted on the basis of their mass/luminosity.

\item The data of the "Illustris" simulation predict BCGs and II-BCGs light profiles quite similar
to the observed ones. The effective radii and the velocity dispersions are also quite similar to that observed, while clusters are still systematically less massive and luminous that real clusters at $z=0$. 
The average light profiles of simulated galaxies can be well superposed to that of real galaxies with small systematic deviations occurring only in the inner and outer regions. These differences are likely originated by a still non perfect receipt adopted for the merging and feedback effects. Simulated clusters have steeper profiles in the center than real clusters.

\item We have used the merger trees of BCGs and II-BCGs following their main progenitor branches to shed light on the evolution of their effective and half-mass radius with time: we have focused in particular on two reference past epochs, namely $z=0.2$ and $z=0.8$. It is evident that II-BCGs are still evolving at the selected epochs, while BCGs have seen only a moderate evolution. The progenitors of II-BCGs were not necessarily the II-BCG of the clusters at these redshifts, while BCGs remain the brightest objects of the clusters. The effective radius can either be larger or smaller with respect to earlier epochs because of merging events.

From the comparison of \re\ at different epochs we argue that:
(i) negative variations can be due to spontaneous dynamical cooling of the innermost regions, to mergers having relatively massive star clumps and effectively reaching the innermost regions of the main progenitor, or to a combination of both factors. Stripping events are also possible;
(ii) positive variations are likely due to the accretion of star clumps having non negligible mass and/or objects where infall is still occurring; here the amplitudes of the variations depend on the mass of the star clump compared to that of the central bulk as well as on its radial position.
Clearly any combination between these situations is possible; for example, massive galaxies most likely acquire smaller subhalos from the outskirts so that slight expansions of central volumes enclosing half the whole light (or mass) are expected. Moreover, because of the temporal separation between the three reference epochs, the galaxies of the samples might acquire large fractions of star (or DM particles) immediately before $z=0.2$ or $z=0$. This means that galaxies are not fully relaxed. 

\item Motivated by the past theoretical studies devoted to the application of the S\'ersic's law to the distribution of matter, we have extended the analysis of simulated structures to the surface mass density profiles ($\mu_m$, in terms of both average and single profiles). We find that the luminous matter plays a minimal role in shaping the profiles and DM gives the chief contributor. However the average profiles are dominated by the former in the central regions of all structures, sensibly increasing the slopes of the profiles for $r/R_{1/2}\lesssim 0.3$. A significant departure from the self-similarity of the average profiles is evident over $r/R_{1/2}\sim 3$ for galaxies and $r/R_{1/2}\sim 1$ for clusters. Above these limits, constant minor mergers cause the structures to never reach relaxation in a Hubble time, while below these limits the S\'ersic's law is able to fit the average profiles with very low rms. 
The profile analysis confirms that the luminous and dark components are strictly affecting each other. 

\end{itemize}

In conclusion galaxy clusters and ETGs show similar light and mass profiles that might come from the same original profile later evolved in different ways. The deviations form this 'seed' profile occur only in the inner and outer regions, where many physical effects are at work, in particular for the baryon component. 
All systems are non-homologous, in the sense that their shapes, represented by the S\'ersic index $n$, vary considerably within the same class of objects.

%\acknowledgments
\section*{Acknowledgments}
We kindly thank the anonymous referee for his/her comments and suggestions that have greatly improved the paper. M.D. thanks Didier Fraix-Burnet for helpful discussions on the text of the paper.
C.C. likes to thank the Department of Physics and Astronomy of the Padua University for the hospitality and computational support.

\appendix
\label{Appendix}

%\section{Tables} \label{sec:tables}

\begin{table}[h]
    \centering
    \caption{Useful physical quantities of 5 random ETGs extracted from Illustris dataset at $z=0$
    (see text for more details).}
    \renewcommand{\arraystretch}{0.8}
    \begin{tabular}{|c|c c c c|}
    \hline
         subhalo ID & cluster ID & $M_{20}$ & $G$ & $M_* (10^{10}M_{\odot})$  \\
         \hline
        000133 &  00  & -1.99 & 0.60 & 0.438672 \\
        051847 &  03  & -1.99 & 0.56 & 0.796838 \\
        080788 &  07  & -1.99 & 0.56 & 0.580556 \\
        086214 &  08  & -2.01 & 0.56 & 0.914614 \\
        120636 &  15  & -1.99 & 0.57 & 0.692473 \\
        \hline
    \end{tabular}
    \label{tab:retg_vals}
\end{table}

\begin{table*}[]
\begin{center}
\caption{Single S\'ersic parameters for surface mass density profiles of BCGs at $z=0$.}
\label{tab_sersic_single_mass_1}
\renewcommand{\arraystretch}{0.8}
\begin{tabular}{|c| c c c c c c c| }
\hline
  %$\mu_m(R_{1/2})$  & $n$  & $\sigma$      &$b(n)$ &  $\sigma$ & species redshift id &$<\Delta\mu_m^2>$  \\
       & $\mu_m(R_{1/2})$  & $n$  & $\Delta n$      & $b(n)$ &  $\Delta b(n)$ & $\sigma_m(\leq R_{1/2})$ & $\sigma_m(\leq3R_{1/2})$  \\
  \hline
BCG@z=0.0,id:0 & 16.06 &  0.95 &  0.04 &  2.94 &  0.16 &  0.30 &  0.21 \\ 
BCG@z=0.0,id:16937 & 15.76 &  0.53 &  0.04 &  1.06 &  0.17 &  0.64 &  0.45 \\ 
BCG@z=0.0,id:30430 & 15.98 &  1.39 &  0.08 &  3.22 &  0.20 &  0.23 &  0.17 \\ 
BCG@z=0.0,id:41088 & 16.51 &  2.61 &  0.11 &  5.33 &  0.21 &  0.03 &  0.08 \\ 
BCG@z=0.0,id:51811 & 16.64 &  1.12 &  0.07 &  3.27 &  0.24 &  0.17 &  0.27 \\ 
BCG@z=0.0,id:59384 & 16.21 &  1.50 &  0.12 &  3.49 &  0.30 &  0.21 &  0.24 \\ 
BCG@z=0.0,id:66080 & 15.66 &  0.81 &  0.07 &  1.94 &  0.25 &  0.52 &  0.40 \\ 
BCG@z=0.0,id:73663 & 17.74 &  4.03 &  0.66 &  8.17 &  1.26 &  0.19 &  0.23 \\ 
BCG@z=0.0,id:80734 & 15.96 &  1.22 &  0.09 &  2.94 &  0.27 &  0.33 &  0.28 \\ 
BCG@z=0.0,id:86186 & 16.30 &  1.68 &  0.07 &  3.72 &  0.17 &  0.15 &  0.12 \\ 
BCG@z=0.0,id:93165 & 16.23 &  2.72 &  0.16 &  5.35 &  0.31 &  0.09 &  0.11 \\ 
BCG@z=0.0,id:99148 & 16.61 &  2.20 &  0.13 &  4.73 &  0.28 &  0.13 &  0.13 \\ 
BCG@z=0.0,id:104798 & 18.16 &  6.27 &  0.88 & 10.57 &  1.40 &  0.07 &  0.11 \\ 
BCG@z=0.0,id:110567 & 17.03 &  2.53 &  0.14 &  5.35 &  0.28 &  0.14 &  0.11 \\ 
BCG@z=0.0,id:114300 & 15.88 &  0.81 &  0.08 &  1.94 &  0.29 &  0.56 &  0.46 \\ 
BCG@z=0.0,id:117343 & 15.95 &  1.44 &  0.11 &  3.10 &  0.27 &  0.26 &  0.23 \\ 
BCG@z=0.0,id:120615 & 16.61 &  1.73 &  0.07 &  3.70 &  0.17 &  0.12 &  0.11 \\ 
BCG@z=0.0,id:123773 & 15.82 &  0.29 &  0.04 &  0.24 &  0.11 &  0.88 &  0.94 \\ 
BCG@z=0.0,id:127228 & 15.90 &  1.15 &  0.07 &  2.67 &  0.21 &  0.34 &  0.23 \\ 
BCG@z=0.0,id:129770 & 16.62 &  1.64 &  0.08 &  3.74 &  0.20 &  0.17 &  0.14 \\ 
\hline
\end{tabular}
\end{center}
\end{table*}

\begin{table*}
\begin{center}
\caption{Single S\'ersic parameters for surface mass density profiles of BCGs at $z=0.2$.}
\label{tab_sersic_single_mass_2}
\renewcommand{\arraystretch}{0.8}
\begin{tabular}{|c| c c c c c c c| }
\hline
  %$\mu_m(R_{1/2})$  & $n$  & $\sigma$      &$b(n)$ &  $\sigma$ & species redshift id &$<\Delta\mu_m^2>$  \\
       & $\mu_m(R_{1/2})$  & $n$  & $\Delta n$      & $b(n)$ &  $\Delta b(n)$ & $\sigma_m(\leq R_{1/2})$ & $\sigma_m(\leq3R_{1/2})$  \\
  \hline
BCG@z=0.2,id:0 & 16.40 &  1.57 &  0.08 &  3.96 &  0.23 &  0.11 &  0.17 \\ 
BCG@z=0.2,id:11777 & 15.39 &  0.80 &  0.05 &  1.90 &  0.19 &  0.47 &  0.31 \\ 
BCG@z=0.2,id:24088 & 16.17 &  1.29 &  0.07 &  3.01 &  0.18 &  0.20 &  0.17 \\ 
BCG@z=0.2,id:33383 & 16.43 &  0.94 &  0.06 &  2.63 &  0.25 &  0.39 &  0.34 \\ 
BCG@z=0.2,id:40441 & 17.60 &  2.42 &  0.10 &  5.42 &  0.23 &  0.05 &  0.09 \\ 
BCG@z=0.2,id:46753 & 15.96 &  0.97 &  0.08 &  2.24 &  0.25 &  0.43 &  0.33 \\ 
BCG@z=0.2,id:52457 & 16.16 &  1.44 &  0.10 &  3.26 &  0.24 &  0.21 &  0.20 \\ 
BCG@z=0.2,id:57650 & 17.26 &  1.63 &  0.20 &  3.62 &  0.47 &  0.35 &  0.33 \\ 
BCG@z=0.2,id:63467 & 16.22 &  2.02 &  0.09 &  4.20 &  0.20 &  0.06 &  0.11 \\ 
BCG@z=0.2,id:68485 & 16.68 &  1.74 &  0.08 &  4.32 &  0.22 &  0.12 &  0.15 \\ 
BCG@z=0.2,id:74066 & 15.98 &  1.02 &  0.06 &  2.48 &  0.20 &  0.37 &  0.25 \\ 
BCG@z=0.2,id:78783 & 16.45 &  1.66 &  0.13 &  3.84 &  0.32 &  0.14 &  0.22 \\ 
BCG@z=0.2,id:82944 & 17.71 &  5.47 &  1.37 & 10.80 &  2.56 &  0.11 &  0.26 \\ 
BCG@z=0.2,id:86310 & 16.92 &  2.42 &  0.16 &  4.76 &  0.31 &  0.13 &  0.13 \\ 
BCG@z=0.2,id:90381 & 16.44 &  1.36 &  0.13 &  2.95 &  0.32 &  0.32 &  0.29 \\ 
BCG@z=0.2,id:93818 & 16.30 &  0.82 &  0.06 &  1.85 &  0.21 &  0.41 &  0.33 \\ 
BCG@z=0.2,id:97243 & 16.34 &  1.31 &  0.09 &  2.85 &  0.24 &  0.24 &  0.22 \\ 
BCG@z=0.2,id:100284 & 16.34 &  1.47 &  0.10 &  3.36 &  0.25 &  0.23 &  0.21 \\ 
BCG@z=0.2,id:103005 & 16.76 &  1.33 &  0.08 &  3.27 &  0.23 &  0.19 &  0.21 \\ 
BCG@z=0.2,id:132611 & 16.42 &  1.38 &  0.09 &  3.03 &  0.22 &  0.23 &  0.20 \\ 
\hline
\end{tabular}
\end{center}
\end{table*}

\begin{table*}
\begin{center}
\caption{Single S\'ersic parameters for surface mass density profiles of BCGs at $z=0.8$.}
\label{tab_sersic_single_mass_3}
\renewcommand{\arraystretch}{0.8}
\begin{tabular}{|c| c c c c c c c| }
\hline
  %$\mu_m(R_{1/2})$  & $n$  & $\sigma$      &$b(n)$ &  $\sigma$ & species redshift id &$<\Delta\mu_m^2>$  \\
       & $\mu_m(R_{1/2})$  & $n$  & $\Delta n$      & $b(n)$ &  $\Delta b(n)$ & $\sigma_m(\leq R_{1/2})$ & $\sigma_m(\leq3R_{1/2})$  \\
  \hline
BCG@z=0.8,id:0 & 16.83 &  1.05 &  0.09 &  2.92 &  0.33 &  0.31 &  0.40 \\ 
BCG@z=0.8,id:5951 & 16.09 &  0.57 &  0.07 &  1.26 &  0.29 &  0.68 &  0.71 \\ 
BCG@z=0.8,id:10747 & 17.00 &  1.41 &  0.10 &  3.33 &  0.26 &  0.29 &  0.22 \\ 
BCG@z=0.8,id:14807 & 12.33 &  1.35 &  0.10 &  2.96 &  0.24 &  0.16 &  0.22 \\ 
BCG@z=0.8,id:18160 & 17.03 &  1.32 &  0.07 &  3.26 &  0.20 &  0.25 &  0.19 \\ 
BCG@z=0.8,id:20984 & 16.97 &  1.05 &  0.08 &  2.49 &  0.26 &  0.34 &  0.32 \\ 
BCG@z=0.8,id:23233 & 17.23 &  1.50 &  0.09 &  3.89 &  0.26 &  0.26 &  0.21 \\ 
BCG@z=0.8,id:25993 & 16.63 &  1.43 &  0.11 &  3.13 &  0.28 &  0.23 &  0.23 \\ 
BCG@z=0.8,id:28932 & 16.48 &  0.85 &  0.06 &  2.08 &  0.23 &  0.44 &  0.35 \\ 
BCG@z=0.8,id:32025 & 16.62 &  0.74 &  0.06 &  1.76 &  0.24 &  0.48 &  0.42 \\ 
BCG@z=0.8,id:34434 & 16.80 &  1.44 &  0.11 &  3.21 &  0.29 &  0.26 &  0.24 \\ 
BCG@z=0.8,id:38876 & 16.93 &  1.29 &  0.09 &  2.83 &  0.24 &  0.27 &  0.23 \\ 
BCG@z=0.8,id:40495 & 16.54 &  1.28 &  0.06 &  2.92 &  0.15 &  0.20 &  0.15 \\ 
BCG@z=0.8,id:42332 & 13.58 &  0.50 &  0.05 &  1.00 &  0.23 &  0.71 &  0.61 \\ 
BCG@z=0.8,id:44152 & 17.30 &  0.83 &  0.05 &  2.36 &  0.22 &  0.39 &  0.34 \\ 
BCG@z=0.8,id:45976 & 16.68 &  1.44 &  0.09 &  3.23 &  0.24 &  0.28 &  0.20 \\ 
BCG@z=0.8,id:60183 & 17.51 &  1.31 &  0.09 &  3.44 &  0.29 &  0.31 &  0.27 \\ 
BCG@z=0.8,id:64890 & 16.77 &  1.39 &  0.14 &  3.31 &  0.38 &  0.36 &  0.33 \\ 
BCG@z=0.8,id:108426 & 17.22 &  1.31 &  0.09 &  3.29 &  0.28 &  0.36 &  0.26 \\ 
BCG@z=0.8,id:134538 & 17.04 &  0.99 &  0.07 &  2.34 &  0.23 &  0.43 &  0.29 \\ 
\hline
\end{tabular}
\end{center}
\end{table*}

\begin{table*}
\begin{center}
\caption{Single S\'ersic parameters for surface mass density profiles of II-BCGs at $z=0$.}
\label{tab_sersic_single_mass_4}
\renewcommand{\arraystretch}{0.8}
\begin{tabular}{|c| c c c c c c c| }
\hline
  %$\mu_m(R_{1/2})$  & $n$  & $\sigma$      &$b(n)$ &  $\sigma$ & species redshift id &$<\Delta\mu_m^2>$  \\
       & $\mu_m(R_{1/2})$  & $n$  & $\Delta n$      & $b(n)$ &  $\Delta b(n)$ & $\sigma_m(\leq R_{1/2})$ & $\sigma_m(\leq3R_{1/2})$  \\
  \hline
SBCG@z=0.0,id:1 & 16.40 &  0.59 &  0.07 &  1.30 &  0.28 &  0.84 &  0.66 \\ 
SBCG@z=0.0,id:16938 & 14.81 &  1.23 &  0.08 &  2.96 &  0.23 &  0.32 &  0.23 \\ 
SBCG@z=0.0,id:30433 & 16.01 &  3.20 &  0.29 &  6.66 &  0.58 &  0.16 &  0.15 \\ 
SBCG@z=0.0,id:41092 & 14.88 &  1.90 &  0.19 &  4.08 &  0.43 &  0.23 &  0.25 \\ 
SBCG@z=0.0,id:51814 & 14.81 &  0.87 &  0.10 &  2.04 &  0.35 &  0.61 &  0.52 \\ 
SBCG@z=0.0,id:59386 & 14.91 &  1.34 &  0.19 &  3.16 &  0.51 &  0.43 &  0.46 \\ 
SBCG@z=0.0,id:66082 & 15.10 &  1.59 &  0.14 &  3.84 &  0.36 &  0.35 &  0.27 \\ 
SBCG@z=0.0,id:73664 & 15.36 &  2.00 &  0.10 &  3.99 &  0.21 &  0.14 &  0.12 \\ 
SBCG@z=0.0,id:80735 & 15.41 &  1.11 &  0.08 &  2.56 &  0.24 &  0.31 &  0.27 \\ 
SBCG@z=0.0,id:86187 & 15.25 &  1.35 &  0.11 &  3.06 &  0.28 &  0.31 &  0.26 \\ 
SBCG@z=0.0,id:93166 & 16.76 &  3.10 &  0.19 &  6.42 &  0.37 &  0.14 &  0.10 \\ 
SBCG@z=0.0,id:99149 & 14.61 &  0.69 &  0.06 &  1.67 &  0.27 &  0.72 &  0.51 \\ 
SBCG@z=0.0,id:104799 & 15.33 &  0.93 &  0.09 &  2.19 &  0.30 &  0.54 &  0.42 \\ 
SBCG@z=0.0,id:110566 & 16.57 &  1.84 &  0.08 &  4.11 &  0.19 &  0.15 &  0.12 \\ 
SBCG@z=0.0,id:114301 & 15.33 &  1.52 &  0.15 &  3.43 &  0.37 &  0.31 &  0.29 \\ 
SBCG@z=0.0,id:117346 & 15.30 &  2.37 &  0.11 &  4.84 &  0.23 &  0.14 &  0.10 \\ 
SBCG@z=0.0,id:120623 & 15.80 &  1.12 &  0.08 &  2.79 &  0.25 &  0.37 &  0.28 \\ 
SBCG@z=0.0,id:123774 & 14.68 &  1.57 &  0.14 &  3.60 &  0.36 &  0.30 &  0.27 \\ 
SBCG@z=0.0,id:127230 & 14.80 &  1.93 &  0.13 &  3.99 &  0.28 &  0.16 &  0.16 \\ 
SBCG@z=0.0,id:129771 & 16.32 &  1.39 &  0.09 &  3.34 &  0.26 &  0.32 &  0.23 \\ 
\hline
\end{tabular}
\end{center}
\end{table*}

\begin{table*}
\begin{center}
\caption{Single S\'ersic parameters for surface mass density profiles of II-BCGs at $z=0.2$.}
\label{tab_sersic_single_mass_5}
\renewcommand{\arraystretch}{0.8}
\begin{tabular}{|c| c c c c c c c| }
\hline
  %$\mu_m(R_{1/2})$  & $n$  & $\sigma$      &$b(n)$ &  $\sigma$ & species redshift id &$<\Delta\mu_m^2>$  \\
       & $\mu_m(R_{1/2})$  & $n$  & $\Delta n$      & $b(n)$ &  $\Delta b(n)$ & $\sigma_m(\leq R_{1/2})$ & $\sigma_m(\leq3R_{1/2})$  \\
  \hline
SBCG@z=0.2,id:11779 & 15.75 &  1.21 &  0.10 &  2.71 &  0.27 &  0.40 &  0.28 \\ 
SBCG@z=0.2,id:24092 & 16.01 &  1.46 &  0.20 &  3.40 &  0.52 &  0.48 &  0.42 \\ 
SBCG@z=0.2,id:33384 & 16.11 &  1.02 &  0.07 &  3.10 &  0.28 &  0.39 &  0.35 \\ 
SBCG@z=0.2,id:40442 & 15.95 &  1.93 &  0.12 &  3.91 &  0.26 &  0.15 &  0.15 \\ 
SBCG@z=0.2,id:46754 & 15.50 &  2.07 &  0.20 &  4.56 &  0.44 &  0.22 &  0.23 \\ 
SBCG@z=0.2,id:57652 & 15.58 &  1.22 &  0.12 &  2.86 &  0.33 &  0.39 &  0.34 \\ 
SBCG@z=0.2,id:68486 & 15.00 &  0.74 &  0.07 &  1.77 &  0.29 &  0.71 &  0.51 \\ 
SBCG@z=0.2,id:74067 & 14.39 &  0.51 &  0.04 &  1.07 &  0.18 &  0.80 &  0.53 \\ 
SBCG@z=0.2,id:82945 & 16.00 &  1.14 &  0.09 &  2.69 &  0.27 &  0.42 &  0.30 \\ 
SBCG@z=0.2,id:86309 & 16.44 &  1.17 &  0.09 &  2.91 &  0.29 &  0.42 &  0.31 \\ 
SBCG@z=0.2,id:90385 & 15.89 &  2.01 &  0.15 &  4.36 &  0.34 &  0.16 &  0.18 \\ 
SBCG@z=0.2,id:97247 & 15.84 &  2.69 &  0.17 &  5.49 &  0.33 &  0.17 &  0.12 \\ 
SBCG@z=0.2,id:100285 & 15.80 &  1.34 &  0.12 &  3.00 &  0.32 &  0.33 &  0.29 \\ 
SBCG@z=0.2,id:103006 & 15.69 &  1.38 &  0.11 &  3.40 &  0.31 &  0.35 &  0.27 \\ 
SBCG@z=0.2,id:105632 & 16.67 &  1.55 &  0.14 &  3.28 &  0.32 &  0.28 &  0.24 \\ 
SBCG@z=0.2,id:137480 & 17.10 &  1.13 &  0.06 &  3.21 &  0.23 &  0.37 &  0.25 \\ 
SBCG@z=0.2,id:176645 & 17.06 &  0.92 &  0.07 &  2.55 &  0.27 &  0.49 &  0.38 \\ 
SBCG@z=0.2,id:180271 & 16.81 &  1.77 &  0.12 &  3.96 &  0.28 &  0.24 &  0.18 \\ 
SBCG@z=0.2,id:230239 & 16.89 &  1.18 &  0.11 &  2.95 &  0.34 &  0.45 &  0.36 \\ 
SBCG@z=0.2,id:358004 & 17.47 &  2.18 &  0.15 &  4.73 &  0.33 &  0.17 &  0.16 \\ 
\hline
\end{tabular}
\end{center}
\end{table*}

\begin{table*}
\begin{center}
\caption{Single S\'ersic parameters for surface mass density profiles of II-BCGs at $z=0.8$.}
\label{tab_sersic_single_mass_6}
\renewcommand{\arraystretch}{0.8}
\begin{tabular}{|c| c c c c c c c| }
\hline
  %$\mu_m(R_{1/2})$  & $n$  & $\sigma$      &$b(n)$ &  $\sigma$ & species redshift id &$<\Delta\mu_m^2>$  \\
       & $\mu_m(R_{1/2})$  & $n$  & $\Delta n$      & $b(n)$ &  $\Delta b(n)$ & $\sigma_m(\leq R_{1/2})$ & $\sigma_m(\leq3R_{1/2})$  \\
  \hline
SBCG@z=0.8,id:36780 & 16.71 &  2.08 &  0.13 &  4.20 &  0.27 &  0.12 &  0.14 \\ 
SBCG@z=0.8,id:53591 & 17.20 &  1.05 &  0.10 &  2.44 &  0.30 &  0.52 &  0.36 \\ 
SBCG@z=0.8,id:72814 & 16.90 &  1.01 &  0.09 &  2.39 &  0.29 &  0.53 &  0.37 \\ 
SBCG@z=0.8,id:75685 & 16.99 &  1.40 &  0.08 &  3.16 &  0.20 &  0.28 &  0.17 \\ 
SBCG@z=0.8,id:83154 & 16.92 &  0.88 &  0.05 &  2.40 &  0.19 &  0.41 &  0.28 \\ 
SBCG@z=0.8,id:88212 & 16.82 &  0.93 &  0.09 &  2.24 &  0.30 &  0.57 &  0.42 \\ 
SBCG@z=0.8,id:92482 & 16.60 &  0.65 &  0.05 &  1.43 &  0.20 &  0.57 &  0.41 \\ 
SBCG@z=0.8,id:101263 & 17.32 &  2.19 &  0.24 &  4.43 &  0.49 &  0.24 &  0.24 \\ 
SBCG@z=0.8,id:106868 & 17.74 &  1.82 &  0.15 &  3.84 &  0.33 &  0.26 &  0.20 \\ 
SBCG@z=0.8,id:114022 & 17.15 &  1.45 &  0.10 &  3.61 &  0.27 &  0.32 &  0.22 \\ 
SBCG@z=0.8,id:118576 & 16.45 &  0.96 &  0.09 &  2.46 &  0.33 &  0.54 &  0.43 \\ 
SBCG@z=0.8,id:125781 & 17.07 &  1.93 &  0.11 &  4.25 &  0.26 &  0.17 &  0.15 \\ 
SBCG@z=0.8,id:128161 & 17.35 &  1.39 &  0.13 &  3.13 &  0.34 &  0.34 &  0.30 \\ 
SBCG@z=0.8,id:171082 & 17.38 &  1.77 &  0.18 &  3.98 &  0.42 &  0.33 &  0.27 \\ 
SBCG@z=0.8,id:174742 & 17.42 &  1.89 &  0.14 &  4.24 &  0.32 &  0.19 &  0.19 \\ 
SBCG@z=0.8,id:175715 & 17.24 &  1.41 &  0.15 &  3.34 &  0.39 &  0.41 &  0.34 \\ 
SBCG@z=0.8,id:202961 & 17.70 &  1.96 &  0.14 &  4.22 &  0.31 &  0.19 &  0.18 \\ 
SBCG@z=0.8,id:213485 & 18.30 &  2.71 &  0.26 &  5.37 &  0.50 &  0.17 &  0.17 \\ 
SBCG@z=0.8,id:236478 & 17.25 &  2.38 &  0.11 &  4.85 &  0.21 &  0.11 &  0.09 \\ 
SBCG@z=0.8,id:281354 & 17.54 &  1.39 &  0.13 &  2.97 &  0.32 &  0.26 &  0.28 \\ 
\hline
\end{tabular}
\end{center}
\end{table*}

\begin{table*}
\begin{center}
\caption{Single S\'ersic parameters for surface mass density profiles of clusters at $z=0$.}
\label{tab_sersic_single_mass_7}
\renewcommand{\arraystretch}{0.8}
\begin{tabular}{|c| c c c c c c c| }
\hline
  %$\mu_m(R_{1/2})$  & $n$  & $\sigma$      &$b(n)$ &  $\sigma$ & species redshift id &$<\Delta\mu_m^2>$  \\
       & $\mu_m(R_{1/2})$  & $n$  & $\Delta n$      & $b(n)$ &  $\Delta b(n)$ & $\sigma_m(\leq R_{1/2})$ & $\sigma_m(\leq3R_{1/2})$  \\
  \hline
CL@z=0.0,id:0 & 17.63 &  0.79 &  0.17 &  3.61 &  0.85 &  0.48 &  0.97 \\ 
CL@z=0.0,id:1 & 15.77 &  0.38 &  0.05 &  0.78 &  0.27 &  0.67 &  0.95 \\ 
CL@z=0.0,id:2 & 16.27 &  0.59 &  0.10 &  1.75 &  0.52 &  0.63 &  1.01 \\ 
CL@z=0.0,id:3 & 17.51 &  0.20 &  0.06 &  0.32 &  0.28 &  1.34 &  1.37 \\ 
CL@z=0.0,id:4 & 14.01 &  4.33 &  0.85 &  3.87 &  0.72 &  0.09 &  0.11 \\ 
CL@z=0.0,id:5 & 16.84 &  3.17 &  1.17 &  6.56 &  2.32 &  0.24 &  0.62 \\ 
CL@z=0.0,id:6 & 16.28 &  0.40 &  0.05 &  1.51 &  0.33 &  0.79 &  0.72 \\ 
CL@z=0.0,id:7 & 19.16 & 11.22 & 14.97 & 17.99 & 23.06 &  0.35 &  0.61 \\ 
CL@z=0.0,id:8 & 15.90 &  3.58 &  0.89 &  5.86 &  1.39 &  0.08 &  0.31 \\ 
CL@z=0.0,id:9 & 16.12 &  0.22 &  0.04 &  0.16 &  0.12 &  1.10 &  1.39 \\ 
CL@z=0.0,id:10 & 14.83 &  3.15 &  0.26 &  4.47 &  0.35 &  0.07 &  0.10 \\ 
CL@z=0.0,id:11 & 16.63 &  0.30 &  0.05 &  0.65 &  0.29 &  1.11 &  1.27 \\ 
CL@z=0.0,id:12 & 17.49 &  0.15 &  0.03 &  0.47 &  0.25 &  0.95 &  0.77 \\ 
CL@z=0.0,id:13 & 17.89 & 15.15 & 48.15 & 19.85 & 61.10 &  0.25 &  0.90 \\ 
CL@z=0.0,id:14 & 15.63 &  1.35 &  0.15 &  2.61 &  0.33 &  0.33 &  0.29 \\ 
CL@z=0.0,id:15 & 16.45 &  0.52 &  0.07 &  1.47 &  0.40 &  0.80 &  0.94 \\ 
CL@z=0.0,id:16 & 15.61 &  2.05 &  0.23 &  3.18 &  0.36 &  0.14 &  0.19 \\ 
CL@z=0.0,id:17 & 14.69 & - & - & - & - & - & - \\ 
CL@z=0.0,id:18 & 15.94 &  1.34 &  0.12 &  2.90 &  0.30 &  0.28 &  0.28 \\ 
CL@z=0.0,id:19 & 18.13 & 18.53 & 36.96 & 30.59 & 59.40 &  0.23 &  0.59 \\ 
\hline
\end{tabular}
\end{center}
\end{table*}

\end{document}